\documentclass{aa}  

\newcommand{\msun}{\mathrm{M}_\odot}
\newcommand{\bq}{\begin{equation}}
\newcommand{\eq}{\end{equation}}
\newcommand{\bqn}{\begin{eqnarray}}
\newcommand{\eqn}{\end{eqnarray}}
\def\PGPU{$\varphi-$GPU }

\usepackage{graphicx}
\usepackage{txfonts}
\usepackage{upgreek}
\usepackage[dvipsnames]{xcolor}
\usepackage{hyperref}
\usepackage[normalem]{ulem}

\hypersetup{
    colorlinks = true,
    linkcolor = blue,
    citecolor = blue,
    filecolor = cyan,
    urlcolor = purple
}

\newcommand{\orcid}[1]{}

\begin{document} 

\title{Dynamical model of Praesepe and its tidal tails}


\author{L. Weis\inst{1}\orcid{0009-0007-4230-5987}
        \and
        C. West\inst{1}
        \and          
        A. Just\inst{1}\fnmsep\thanks{email: just@ari.uni-heidelberg.de}\orcid{0000-0002-5144-9233}
        \and
        P. Berczik\inst{2,3,4}\orcid{0000-0003-4176-152X}
        \and        M.~Ishchenko\inst{4,3,2}\orcid{0000-0002-6961-8170}
        \and
        S. R\"oser\inst{1}
        \and          
        E. Schilbach\inst{1}
        \and
        B.~Shukirgaliyev\inst{5,6,7}\orcid{0000-0002-4601-7065}
       }

\institute{Zentrum f\"ur Astronomie der Universit\"at Heidelberg, Astronomisches Rechen-Institut, M\"{o}nchhofstra\ss{}e 12-14, 69120 Heidelberg, Germany
\and
Nicolaus Copernicus Astronomical Centre Polish Academy of Sciences, ul. Bartycka 18, 00-716 Warsaw, Poland
\and
Fesenkov Astrophysical Institute, Observatory 23, 050020 Almaty, Kazakhstan
\and
Main Astronomical Observatory, National Academy of Sciences of Ukraine, 27 Akademika Zabolotnoho St., 03143 Kyiv, Ukraine
\and
Physics Department, Nazarbayev University, 53 Kabanbay Batyr Ave. 010000 Astana, Kazakhstan
\and
Heriot-Watt International Faculty, K.~Zhubanov Aktobe Regional University, 263 Zhubanov Brothers str., 030000 Aktobe, Kazakhstan
\and 
Energetic Cosmos Laboratory, Nazarbayev University, 53 Kabanbay Batyr Ave. 010000 Astana, Kazakhstan
}

\date{Received ...; accepted ...}

 
  \abstract
   {The dynamical evolution of open clusters in the tidal field of the Milky Way and the feeding of the disc field star population depend strongly on the initial conditions at the time of gas removal. Detailed dynamical models tailored to individual clusters help us understand the role of open clusters in the Galactic disc evolution.}
   {We present a detailed dynamical model of Praesepe, which reproduces the mass profile, the stellar mass function, and the mass segregation observed with the help of Gaia EDR3 data. Based on this model, we investigate the kinematic properties of the tidal tail stars in detail.}
   {We used direct $N$-body simulations along the eccentric orbit of Praesepe in the tidal field of the Milky Way, where each particle represents one star. The initial mass and size of the cluster, the dynamical state, and the initial mass function were adapted to reach the best-fitting model. Based on this model and a comparison model on a circular orbit, we analysed the stars in the tidal tails in terms of 
   density, angular momentum, and orbit shapes.
   }
   {Praesepe can be well reproduced by a cluster model with concentrated star formation in a supervirial state after instantaneous gas expulsion,
   adopting a global star formation efficiency of 17\%. However, the mass distribution inside the cluster is highly sensitive to the random initialisation. About 75\% of the initially 7500\,$\msun$ are lost in the violent relaxation phase, and the observed mass segregation can be understood by two-body relaxation. The tidal tails have a length of about $1\,\mathrm{kpc}$ and show a vertical oscillation along the cluster centre's orbit, which leads to temporal asymmetries in the tidal tails vertical thickness. As a major result, we find that the self-gravity of the tail stars is the dominant force altering the angular momentum of the tail stars. For a typical star, the total change after escaping is about $1.6 \, \mathrm{kpc \, km \, s^{-1}}$. This corresponds to an offset in guiding radius of $7 \, \mathrm{pc}$, where tail stars contribute up to $70 \%$ to the alteration. Additionally, the epicyclic motion leads to an increasing width of the tidal tails. The total radial shift 
   of the orbit of the cluster in the Galactic plane can exceed 50 pc. This effect is not a result of the eccentricity of the orbit.}
   {}

   \maketitle

\section{Introduction}

Star clusters lose their members in the tidal field of the host galaxy. Most escaped stars have orbital parameters similar to those of the parent cluster. Thus, they form leading and trailing tidal tails along the cluster's orbit. 

Understanding the structure of tidal tails can help us uncover more information about the Galactic potential and star formation history. An analysis of the structure of these tidal tails using numerical simulations has been performed with star clusters on circular orbits in a point mass galaxy by \cite{Capuzzo-Dolcetta2005, Montuori2007, kupper2008structure}, in the Milky Way by \cite{just2009quantitative}, and on both eccentric and circular orbits by \cite{kupper2010tidal, kupper2012more}. These tidal tails have clumps or over- and under-densities, usually attributed to the influence of gravitational shocks or passages through a Galactic disc or motion in a triaxial potential. \cite{kupper2008structure} gave an epicyclic theory to explain the formation of clumps in the case of a constant tidal field. \cite{just2009quantitative} presented the epicyclic theory in a more general context. Essentially, clumps arise due to the epicyclic motion of the stars in time-dependent as well as stationary tidal fields.

The tidal tails were observed for many globular clusters \citep[e.g. ][ and many others]{odenkirchen2001detection, Belokurov+2006, GrillmairDionatos2006,Grillmair2009,Juric+2008, Lane+2012}. If the location of the globular clusters is outside the Galactic disc, the stellar overdensities created by tidal tails can be found if the field star density around the tails is very low. However, finding the tidal tails of open clusters (OC) was difficult due to their location in the Galactic disc, where the sky is strongly dominated by field stars. Only with the high-precision observational data provided by Gaia \citep{gaia2018gaia,Brown2021} did it become possible to see some evidence of tidal tails for nearby OCs. Up to date tidal tails were identified for a few OCs only, i.e. Hyades \citep{roser2019hyades, MeingastAlves2019}, Praesepe \citep{roser2019praesepe}, Coma Berenices \citep{Tang+2019}, and a few other clusters \citep{Gao2020, Bhattacharya+2021, Ye+2021}. In most cases, the classical convergence point (CP) method, in which the projection effect of a common space velocity vector at the sky is taken properly into account, has been used to find cluster members including nearby tail stars. For extended tidal tails, the gradient in space velocity along the tails needs to be taken into account. As a first step, it is helpful to perform tailored numerical simulations to predict their observable quantities.  In the past, there were only a few such simulations \citep[e.g.][]{Chumak+2010,ernst+2011,Kovaleva+2020}. In a second crucial step, the predicted trends in velocity need to be taken into account in order to find tidal tail members at larger distances from the cluster. In \citet{Jerabkova2021}, the so-called compact CP (CCP) method was developed, which takes into account a linear trend in space velocity as a function of distance to the cluster. The authors applied the new method with Gaia DR3 data to the Hyades cluster and identify tidal tails to a distance of about 500 pc. \citet{Boffin2022} applied the CCP method to the OC NGC 752 and find tidal tail out to a distance of more than 130\,pc.

In this work, we present a new way to study the tidal tails of OCs with tailored $N$-body simulations. As an example, we consider the Praesepe cluster, one of the nearest OCs consisting of about a thousand stars.

We used cluster data from Gaia EDR3 \citep{Brown2021} out to a distance of 30\,pc identified with the classical CP method and compared the observational data with numerical simulations with respect to the cumulative mass profile, the present-day mass function (PDMF), and the mass segregation. Our initial conditions for $N$-body simulations were based on the model of \citet{parmentier2013local} using a density-driven local star formation efficiency. The star formation efficiency per free fall time, $\epsilon_\mathrm{ff}$, and the duration of the star formation period, $t_\mathrm{SF}$, were assumed to be constant over the whole star forming gas cloud. As a result, the fraction of gas converted to stars is larger in the dense cloud centre compared to the outskirts. The global star formation efficiency, SFE$_\mathrm{gl}$, was determined by the density profile of the gas and the product of $\epsilon_\mathrm{ff}$ and $t_\mathrm{SF}$. Based on this concentrated star formation scenario, it has been shown that a relatively low SFE$_\mathrm{gl} > 15\%$ \citep{shukirgaliyev2017impact} and also relatively high filling factors, $\lambda>0.1$, \citep[see also][]{Shukirgaliyev2018,bek+2019} are sufficient for clusters to survive for hundreds of megayears.

Here, we define the filling factor as the Roche volume filling factor as in \citet{ErnstJust2013} \citep[see also][]{Ernst+2015}; that is, the filling factor, $\lambda$, is the ratio of half-mass radius, $r_\mathrm{h}$, to Jacobi radius, $r_\mathrm{J}$: 
\begin{equation}
 \lambda=\displaystyle \frac{r_\mathrm{h}}{r_\mathrm{J}}.   
\end{equation}
This type of concentrated star formation combined with instantaneous gas expulsion of the remaining molecular cloud leads to a large fraction of unbound  stars leaving the cluster in the violent relaxation phase. The properties of these stars were investigated for the first time in \citet{Dinnbier2020a,Dinnbier2020b} and applied to estimate the ages of young clusters in \citet{Dinnbier2022}.

We selected the best initial conditions and performed a series of 15 random realisations of this model to measure the scatter of cluster properties at present time and to select the best-fit simulation run (BFSR). For this BFSR, we analysed the properties of the tidal tails with respect to the density fluctuations and for the first time with respect to angular momentum variations.

In section \ref{sec:methods}, we describe the basic methods we used for our research. Section \ref{sec:obs_data} describes the observational data we used to perform the tailored $N$-body simulations. We explain how we compared our numerical simulations to observational data in section \ref{sec:best_fit}. We present our main results on tidal tails in section \ref{sec:results}. Then, we summarise our research and make conclusions in section \ref{sec:conc}.

\section{Methods}\label{sec:methods}

\subsection{Jacobi integral and Jacobi radius}\label{sec:jacobi_radius}

The only known constant of motion in star clusters on a circular orbit is the Jacobi integral, $E_\text{J}$, \citep{binney2008galactic}, also known as the Jacobi constant. For a star with velocity, $v$, in the corotating rest frame, it is given by
\begin{equation}
E_\text{J} = \Phi_\text{eff} + \frac{v^2}{2}.
\end{equation}

The effective potential, $\Phi_\text{eff}$, as shown in Eq. (9) in \cite{just2009quantitative}, has saddle points at the inner and outer Lagrange points L1/L2, where stars can leave the cluster if surpassing a critical value, $E_\text{J,crit} = \Phi_\text{eff}(\text{L1/L2})$.
However, there is no clear criterion that determines whether a star is bound or not bound in the cluster. Stars at the Lagrange points below the critical Jacobi energy can only escape if the interaction with other stars changes $E_\text{J}$ above the critical value.

Certainly, stars in the tidal tails with a sufficient distance from the cluster are no longer bound. However, all stars in between are potential escapers, each taking different amounts of time until they reach the Lagrange point or return to the cluster from the outside. Therefore, we used a limiting radius as a pragmatic definition of members in the star cluster, independent of whether the stars are bound or not. This limiting radius is the Jacobi radius, $r_\text{J}$, and is given by
\begin{equation}\label{eq:jacobi_radius}
    r_\text{J} = \left( \frac{GM}{4\Omega^2-\kappa^2}\right)^{\tfrac{1}{3}} = \left( \frac{GM}{(4-\beta^2)\Omega^2}\right)^{\tfrac{1}{3}},
\end{equation}
where $\Omega$, $G$, $M$, $\kappa$, and $\beta = \kappa/\Omega$ denote the angular velocity, gravitational constant, mass of the cluster, epicyclic frequency, and normalised epicyclic frequency, respectively.

In addition, we defined the boundary at which we consider stars to have escaped and become part of the tidal tails to be $2r_\text{J}$, which \cite{ross1997escape} discussed as a non-sharp limit escape criterion.

To summarise, stars with a distance of $r \leq r_\text{J}$ from the cluster centre were treated as bound in the cluster, while stars with $r_\text{J} < r \leq 2r_\text{J}$ were considered to be in an intermediate state. Since our simulations began with supervirial clusters, a large fraction of stars was lost in the violent relaxation phase with a duration of $20\,\mathrm{Myr}$ (see Sec. \ref{sec:models}). All stars with $r > 2r_\text{J}$ at a cluster age of $20\, \mathrm{Myr}$ are called 'escapers', and stars that cross the boundary of $r > 2r_\text{J}$ later are part of the classical tidal tails.

\subsection{Guiding radius}\label{sec:guid_rad}
The equations of motion in an axisymmetric potential $\Phi (R,z)$, which is assumed to be symmetric around the plane $z=0$ \citep{binney2008galactic}, are given in Galactocentric cylindrical coordinates $(R,\phi,z)$ by
\begin{equation}
    \ddot{R} = - \frac{\partial \Phi_\text{eff}}{\partial R} \quad ; \quad \frac{\text{d}}{\text{d}t}(R^2\phi) = 0 \quad ; \quad \ddot{z} = - \frac{\partial \Phi_\text{eff}}{\partial z},
\end{equation}
with $\Phi_\text{eff}$ as the effective potential defined as
\begin{equation}
    \Phi_\text{eff} \equiv \Phi(R,z) + \frac{L^2}{2R^2}.
\end{equation}
Thereby, $L=R^2\dot\phi$ denotes the $z$-component of the (specific) angular momentum.
We find the minima of the effective potential to be $z=0$, due to the assumed symmetry, and in radial direction we find the relation
\begin{equation}\label{eq:guiding_radius}
    \left. \frac{\partial \Phi}{\partial R} \right|_{(R_\mathrm{g}, 0)} = \frac{L^2}{R_\mathrm{g}^3},
\end{equation}
where $R_\mathrm{g}$ is called the guiding radius and the corresponding angular speed $\Omega_\mathrm{g}$ is $\Omega_\mathrm{g} = L/R_\mathrm{g}^2$. As a result, Eq. (\ref{eq:guiding_radius}) describes the condition for a circular orbit at radius $R_\mathrm{g}$ with angular momentum $L$.

\subsection{Models of the open cluster}\label{sec:models}

We used the stellar density profile $\rho_\star(r)$ created with a Plummer model \citep{Plummer1911}. To obtain the BFSR, we created clusters of different sets of initial parameters, namely initial mass ($M$), initial mass function (IMF), filling factor ($\lambda$), and global star formation efficiency (SFE$_\mathrm{gl}$) \citep{shukirgaliyev2017impact}. 

We used a broken power law IMF in a mass range from $0.08\,\msun$ to $100\,\msun$ starting with a Kroupa-IMF \citep{Kroupa2001}. The slopes and the positions of the break points were adapted in order to reproduce the PDMF at the end of the simulation. 
We assumed that our model clusters had a solar metallicity of $Z_\odot = 0.02$ \citep{Grevesse1998}. The results were compared to observational data (Sec. \ref{sec:obs_data}) using the cumulative mass-profile, mass segregation, and the PDMF and were fitted by eye. 
The filling factor was varied between 0.05 and 0.15, and the SFE$_\mathrm{gl}$ between 15\% and 25\%. The initial mass and the parameters of the IMF were freely adapted until the observed bound mass and PDMF was reproduced.
We tested the results on stability by performing the simulations with different random realisations of the same set of initial parameters.

Initial binaries were not taken into account because a binary fraction exceeding 10\% would lead to a very long integration time with the $N$-body code \PGPU used for the simulations. \PGPU resolves the orbits of binaries properly, but since there is no special routine such as regularisation to accelerate the binary orbit integration, it is not feasible for the large number of simulations necessary to find the best model. As already shown in \cite{kroupa1995binaries} for the general dynamical evolution of the cluster, the effect of initial binaries among late type stars would be very small. A similar result was also obtained for the Hyades cluster in \citet{ernst+2011}, which has a similar mass and age as Praesepe. But including binaries may have a significant impact on the IMF and the mass segregation.

\subsection{Tail analysis}\label{sec:tails_analysis}

We analysed the tidal tails of the BFSR run for their structure and dynamics. To eliminate the curvature of the tails in the Cartesian coordinate system, we used a local cylindrical coordinate system as defined as
\begin{equation}\label{eq:loc_cyl}
(X_\text{cyl,loc}, \, Y_\text{cyl,loc}, \, Z_\mathrm{cyl,loc}) = (R-R_\text{c}, \, (\phi- \phi_\text{c}) R_{\text{c}}, \, Z-Z_\mathrm{c}),
\end{equation}
with $R_\text{c}$, $\phi_\text{c}$, and $Z_\mathrm{c}$ being the radius, the phase, and the vertical position of the cluster centre, respectively.

To identify clumps, we analysed the 1D number density along the tails using a tail line through the cluster centre along the tails in the local cylindrical system. For the BFSR, we find an angle of $-3.09^\circ$ between the tails and the horizontal line at $X_\mathrm{cyl,loc} = 0$ (see Fig. \ref{fig:average_tail} (top panel)). The stars were counted in bins along the tail line with a bin width of $25 \, \mathrm{pc}$ to obtain the 1D number density. Additionally, the 3D mass density was also used as a tool to identify clumps  \citep[e.g.][]{just2009quantitative}. We used the standard Smoothed Particle Hydrodynamic (SPH) algorithm \citep{Ber1999} to calculate the 3D stellar density at each point. For that, we calculated the fixed nearest-neighbour numbers for each star and using the current masses of the stars with their individual smoothing lengths, we calculated the smoothed 3D mass density for each selected star.

Since the simulation outcome is dependent on the random seeds (see Sec. \ref{sec:models}), we provide an average 1D number density over all 15 simulations. Hereby, the tail line is determined individually for each simulation. Additionally, we present the average tail shape, which is determined via the radial and vertical extension of the stars per bin using a bin width of $25\,\mathrm{pc}$.

A star in a star cluster can experience three essential phases. The first is when the star is still inside the cluster and gravitationally bound. After that, the star may become a potential escaper through the Lagrange points if its Jacobi energy is raised above the critical value by interaction with other stars inside the cluster or by a decreasing depth of the potential well of the cluster due to expansion or mass loss. The star then escapes in-between $1r_\mathrm{J}$ and $2r_\mathrm{J}$, where only the Jacobi energy is conserved. Once a star has escaped the cluster after surpassing $2r_\mathrm{J}$ (see Sec. \ref{sec:jacobi_radius}), it is part of the tidal tails and travels farther away from the cluster. During this phase in the tail, the energy and the angular momentum are usually assumed to be conserved once the cluster potential no longer significantly influences the star.

We study the dynamics of the stars in the tidal tails using the $z$-component of the specific angular momentum calculated in the Galactic coordinate system, as it is a quantity of conservation, if the interaction with other cluster stars is negligible. The $x$- and $y$-components of the angular momentum are not part of the analysis.

Thereby, we look at the evolution of the angular momentum after a star has escaped the cluster, which is given by 
\begin{equation}\label{eq:DeltaL}
    \Delta L = L_* - L_*(t_\mathrm{out}), 
\end{equation} 
where $L_*$ is the $z$-component of the angular momentum of the star and $L_\mathrm{*}(t_\mathrm{out})$ is the $z$-component of the angular momentum of the star at the time of escape $t_\mathrm{out}$, which is given when a star surpasses $2r_\mathrm{J}$. The angular momentum of a star is determined via the cross product of velocity vector $\boldsymbol{v}$ and position vector $\boldsymbol{r}$ given in the Galactic coordinate system.
Additionally, we define the $z$-component of the angular momentum at the time of escape $t_\mathrm{out}$ relative to the $z$-component of the angular momentum of the cluster centre $ L_\mathrm{c}$ as
\begin{equation}\label{eq:dL_out}
    \delta L_\mathrm{out} = L_*(t_\mathrm{out}) - L_\mathrm{c}.
\end{equation}
Thereby, $L_\mathrm{c}$ is constant 
at all times with negligible numerical fluctuations.

After the escape, as we show in Sec. \ref{sec:angmom}, $\Delta L$ of a star far out in the tail is changed by the specific torque, $\tau$, exerted onto the star by the gravitational force of the cluster stars from different subsystems: stars in the tidal tails, stars inside the cluster, stars in between the cluster and tails, and the stars lost during the violent relaxation phase.

The specific torque, $\tau$, of a star is calculated via the cross-product of its position vector, $\boldsymbol{r}$, and acceleration vector, $\boldsymbol{a}$, in the Galactic coordinate system. The acceleration vector, $\boldsymbol{a}_i$, of a star, $i$, by other stars is given by:
\begin{equation}
    \boldsymbol{a}_i = G \sum_{i\neq j} m_j \frac{\boldsymbol{r}_j - \boldsymbol{r}_i}{|\boldsymbol{r}_j-\boldsymbol{r}_i|^3},
\end{equation}
with $m_j$ and $\boldsymbol{r}_j$ being the mass and the position vector of other stars in the system or chosen subsystems.
We use the $z$-component of the specific torque vector in a discrete time integral to determine the produced angular momentum of $\Delta L_i$ of a star, $i$,
\begin{equation}\label{eq:torque}
    \Delta L_i = \sum_{k=0}^{T} \tau_{i,k} \delta t,
\end{equation}
with $T$ denoting the number of snapshots we integrate over, $\tau_{i,k}$ the determined $z$-component of the star at snapshot $k$, and $\delta t = 0.673 \, \mathrm{Myr}$ the time difference between subsequent snapshots. Note here that this is not the temporal resolution of the simulation, but the data output frequency.

\subsection{Numerical method}\label{sec:num_sim}

In the first step, we integrated the orbit of the cluster centre in the Galactic potential back in time to find the initial position and velocity of the cluster. In the second step, star-by-star random realisations of the cluster stars were integrated forwards in time. For the force calculation, the pairwise forces between all stars born in the cluster were taken into account. Additionally, the background force of the Milky Way in form of an axisymmetric three-component model was taken into account. For this dynamical orbital integration including the stellar evolution, we used the high-order parallel $N$-body code \PGPU\footnote{$N$-body code \PGPU: \\~\url{ https://github.com/berczik/phi-GPU-mole}} which is based on the fourth-order Hermite integration scheme with hierarchical individual block time steps \citep{Berczik2011, BSW2013}. The current version of the \PGPU code uses native GPU support and direct code access to the GPU using the NVIDIA native CUDA library. The present code has been well tested and has already been used to obtain important results in our previous OC simulations \cite{shukirgaliyev2017impact, Shukirgaliyev2018, Bek2021}. 

The Milky Way is represented by an axisymmetric {\tt fixed} three-component potential for disc, bulge, and halo. Similarly to \cite{Kharchenko2009} and \cite{just2009quantitative}, we used the following axisymmetric three-component (bulge-disc-halo) Plummer-Kuzmin model \citep{Miyamoto1975}: 
\begin{equation}
\Phi_{\rm gal}(R,Z) = - \frac{G \cdot M_{\rm b,d,h}}{\sqrt{R^{2}+\Bigl(a_{\rm b,d,h}+\sqrt{Z^{2}+b^{2}_{\rm b,d,h}}\Bigr)^{2}}}, 
\end{equation}
where 
$R=\sqrt{X^{2}+Y^{2}}$ is the planar Galactocentric radius, 
$Z$ is the distance above the plane of the disc, 
$a_{\rm b,d,h}$ are the bulge, disc, and halo scale lengths, 
$b_{\rm b,d,h}$ are the bulge, disc, and halo scale heights, and
$M_{\rm b,d,h}$ are masses of the bulge, disc, and halo, respectively. The exact values of these parameters are presented in Table \ref{tab:pot}. 
It is worthwhile to mention that the $z$-component of angular momentum of a star moving in the axisymmetric Milky Way potential is conserved with high accuracy.

\begin{table}[htbp!]
\caption{Parameters of the fixed axisymmetric three-component Galactic  potential.}
\centering
\begin{tabular}{lrcc}
\hline
\hline
\multicolumn{1}{c}{Galaxy component} & M in M$_\odot$ & a in kpc & b in kpc \\
\hline
\hline
Bulge & $1.4 \times 10^{10}$ & 0.0 & 0.3 \\
Disc  & $9.0 \times 10^{10}$ & 3.3 & 0.3 \\
Halo  & $7.3 \times 10^{11}$ & 0.0 & 25.0 \\
\hline
\end{tabular}
\label{tab:pot}
\end{table} 

The current code also takes into account up-to-date stellar evolution models. The most important updates were made in the stellar evolution part of the code, such as updated metallicity-dependent stellar winds; updated metallicity-dependent core-collapse supernovae, new fallback prescription, and their remnant masses; updated electron-capture supernovae accretion-induced collapse and merger-induced collapse; remnant masses and natal kicks; and BH natal spins and other updates. For more details on the new stellar evolution library, see \cite{Banerjee2020} and \cite{Kamlah2022}.

\section{Observational data}
\label{sec:obs_data}
The samples of candidate members of Praesepe  were selected from the observations in Gaia EDR3 \citep{Brown2021}. The selection process is based on the convergent point (CP) method described, for example, in \citet{roser2019hyades}. Using the space coordinates of the approximate centre of Praesepe, we first selected the subset of all stars from Gaia EDR3 within a radius of $30\, \mathrm{pc}$ around the centre of the cluster. As a next step, from the mean motion of the cluster centre, tangential velocities, $V_{\parallel pred}$ and $V_{\bot pred}$, (parallel and perpendicular to the direction to the CP) were predicted for each star depending on its position on the sky. Then, the predicted tangential velocities were compared with the observed tangential velocities from the Gaia EDR3 measurements. For stars with the exact same motion as the centre, the differences $|\Delta V_{\parallel}|$ and  $|\Delta V_{\bot}|$ would be exactly zero. However, due to the internal velocity dispersion and the uncertainties in the measurements, the observations scatter around the zero point. Stars were counted as candidate members if their $|\Delta V_{\parallel}|$ and $|\Delta V_{\bot}|$ were smaller than three times the standard deviation of their velocity distribution. For Praesepe, these limits were $3.6 \, \mathrm{km} \, \mathrm{s^{-1}}$ for $\Delta V_{\parallel}$  and $1.8 \, \mathrm{km} \, \mathrm{s^{-1}}$ for $\Delta V_{\bot}$. 
So far, this selection may still include observations of low-quality astrometric and/or photometric measurements. For quality checks, \citet{Brown2021} and \citet{Riello2021} proposed using the re-normalised unit-weight-error (RUWE) and the parameter, C*, derived from the value of the 'BP/RP excess factor' in Gaia EDR3. Stars that we selected obey the following criteria: |C*| $<$ 0.1 or  0.1 $\leq$ |C*| $<$ 0.5 and RUWE $\leq$ 2.0. These selection criteria define the astrometric sample. To determine individual masses, we followed the method as used by \citet{roser2019praesepe}. We adopted Parsec 1.2S isochrones \citep{Girardi2002,marigo2008,bressan2012} with [M/H] = +0.15, lg age [Myr] = 8.85 and E(B - V) = 0.027 mag and used the Gaia EDR3 passbands as given in \citet{Riello2021}. The masses were estimated roughly via a mass-luminosity relation in the three Gaia photometric bands, neglecting binary issues. This led to a photometric-based system mass function, since the binary systems were not resolved such that the masses could correspond to the masses derived from the combined photometry of unresolved binaries. The final sample contains 1308 stars with a completeness limit at 17\,mag in G band corresponding to a stellar mass of $0.35 \, \msun$.

\subsection{Cluster properties}\label{sec:clus_prop}

From the observational data, we derived the average cluster position and velocity using all the stars inside the Jacobi radius. For the velocity average, only stars with a measured velocity along the line of sight (VLOS) were considered to obtain a 3D velocity vector. The transformation of the coordinates follows \citet{Johnson1987}. For the position of the North Galactic Pole (NGP), the constrained value from \citet{Karim2016} was used. 

A total of 935 stars were averaged, of which 187 had VLOS, resulting in a cluster velocity relative to the Sun of $V_{\mathrm{c},\odot} =(-43.3 	,	-20.8 	,	-9.0)\,\mathrm{ km\,s^{-1}}$.
For the transformation to the LSR and the Galactocentric coordinate system, we used the peculiar velocity of the Sun relative to the LSR given by \cite{Schoenrich2010}:
\begin{equation*}
(U_\odot,  V_\odot, W_\odot) = (11.1, 12.24 , 7.25 ) \,\mathrm{km\,s^{-1}}
\end{equation*}
leading to
\begin{align*}
    \vec{R}_\text{c,LSR}&=(X_\text{c,LSR},Y_\text{c,LSR},Z_\text{c,LSR}) = (-139.8,-67.7,99.3) \,\mathrm{pc},\\
    \vec{V}_\text{c,LSR} &=(U_\text{c,LSR}, V_\text{c,LSR}, W_\text{c,LSR})=(-32.2, -8.6,-1.8)\,\mathrm{ km\,s^{-1}},
\end{align*}
relative to the LSR.

The calculated coordinates deviate about $2 \, \mathrm{pc}$ and $0.5 \, \mathrm{km\,s^{-1}}$ from those determined by \citet{roser2019praesepe} using data from Gaia DR2 \citep{Brown2018}.  \cite{Babusiaux2018} determined the age of Praesepe as $708_{-91}^{+143} \, \mathrm{Myr}$. We thus adopted the age of $708\, \mathrm{Myr}$ for our simulations. 

\subsection{Cluster orbit}\label{sec:clus_orb}

In the first step, we integrated a point mass in the Galactic potential backwards in time for $708\,\mathrm{Myr}$ in order to determine the initial position and velocity of the cluster. To this end the LSR was positioned in the Galactic model.
In the Galactic coordinate system, the radial distance to the Galactic centre from the Sun is taken as $8.178\, \mathrm{kpc}$ from \citet{Abuter2019}. The proper motion from SgrA$^{*}$ seen from the Sun is $(6.379 \pm 0.026) \, \mathrm{mas\, yr^{-1}}$ \citep{Reid2004}, which can then be converted to a velocity of  $V_{\text{LSR}}+V_\odot=247.30 \, \mathrm{km\,s^{-1}}$. The $Z$ coordinate of the Sun $Z_\odot=(20.8\pm13) \, \mathrm{pc}$ is taken from \citet{Bennett2018}. Hence, the present-day position and velocity of the cluster centre in the Galactocentric Cartesian coordinate system are the following:
 \begin{align*}
    \vec{R}_\mathrm{c}&=(X_\mathrm{c},Y_\mathrm{c},Z_\mathrm{c}) = (-8317.8, -67.7, 	120.1) \,\mathrm{pc},\\
    \vec{V}_\mathrm{c} &=(U_\mathrm{c}, V_\mathrm{c}, W_\mathrm{c})=(-32.2, 226.2, -1.8 )\, \mathrm{ km\,s^{-1}}.
\end{align*}


\section{Parameter study of the initial conditions}\label{sec:best_fit}

The BFSR is found by changing the initial conditions of a simulation run, which is integrated to the present-day, to match the observed properties of stars that exceed the completeness limit. These features include the cumulative mass profile, the mass segregation, and the PDMF of Praesepe inside the Jacobi radius.

The goodness of the fit is assessed visually by iteratively changing the initial parameters. First, different SFE$_\mathrm{gl}$ are tested, affecting the required initial mass due to the increased mass loss in the violent relaxation phase for lower SFE$_\mathrm{gl}$. The filling factor, $\lambda$, is then adjusted, with smaller initial values leading to smaller values at the end of the simulation. The adjustment of the IMF does not significantly change the other properties. Due to the different stellar evolution of light and heavy stars, changing the IMF alters the mass evolution of the cluster and must be taken into account.

We generated our initial star cluster models (coordinates plus velocities and masses) in two separate steps using two different codes. The first is our own developed python code ({\tt plummer-sc.py}) used for the SFE$_\mathrm{gl}$ (gas plus stars) Plummer equilibrium model \citep[see eq. 7-10 and Appendix A,][]{Bek2017}\footnote{\url{https://github.com/berczik/plummer-sc}}. This code is available from the authors' GitHub and is based on the recently very popular python-based {\tt agama} library \citep{agama2019}. 
The second code, which we used for the individual initial mass (IMF) generation of the star cluster is an updated {\tt McLuster}\footnote{\url{ https://github.com/agostinolev/mcluster}} code \citep[see Appendix A,][]{Kuepper2011,mcluster2022}.

Because in the first step we generated the equal mass SFE$_\mathrm{gl}$ Plummer model (with equilibrium gas plus stars system) inside this code, we used the first random generator to get the different sequences of coordinates and velocities with one fixed IMF. During the second step (using the new updated {\tt McLuster} code), we generated the IMF for our cluster. Here we used a second random number to make a variation of the IMF sample for the fixed star cluster positions and velocity data. 

This is why after we found a best-fit model (with a specific set of IMF, positions and velocity) for the first set, we fixed the IMF and then created seven randomisations of the star's positions and velocities inside the cluster.  For the second random set, we fixed the best-fit model initial star's positions and velocities and created seven randomisations of the physically same IMF. So, in the end, we had in total 15 models. One as a best-fit model (after some extended star cluster initial parameter search) plus seven models with the same positions and velocities but different IMF and additional seven models with the same IMF but different positions and velocities.

After identifying a promising set of fitting parameters, we extended our simulations by introducing different random realisations in initial positions and velocities or in the IMF. The sample consists of the BFSR and 14 additional simulations, seven where the random seed for the coordinate and velocity distribution is varied, and seven where the seed for the IMF was varied. Among these 15 random realisations, we identified the most favourable scenario and analysed the collective results of the simulations to derive both an average representation and the scatter of the results.

\subsection{General properties}

\begin{figure}
    \resizebox{\hsize}{!}{\includegraphics{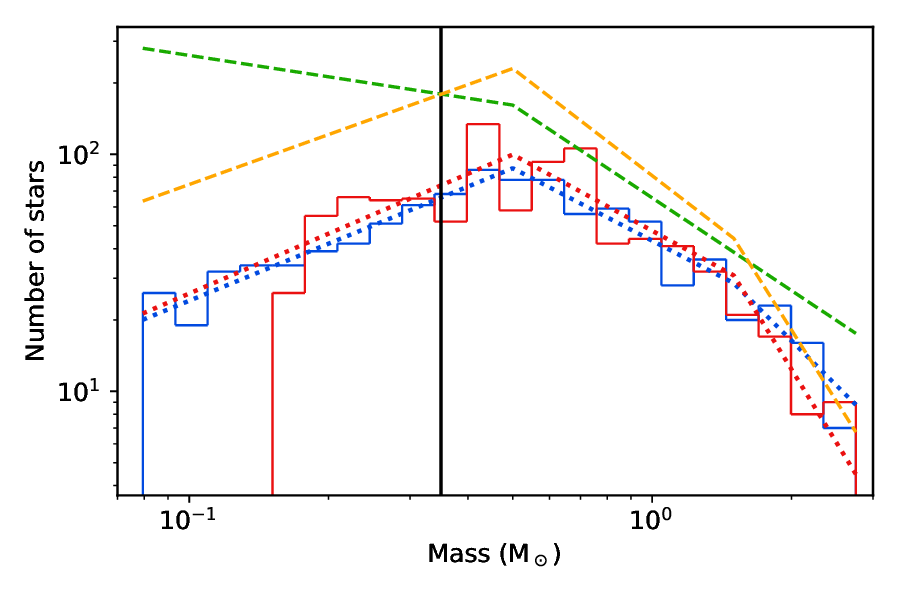}}
    \caption{Mass functions of the BFSR (solid blue histogram) and the observed cluster (solid red histogram) together with a fitted three-part power law (dotted lines in blue and red). Further, the course of the Kroupa IMF \citep{Kroupa2001} (dashed green) and the used IMF for the BFSR (dashed orange) are shown with an arbitrary scaling for good visibility. The completeness limit is marked by the black line. We emphasise that unresolved binary systems in the real cluster have not been taken into account in the models. This can lead to a significant bias \citep{Kroupa1991, Wirth2024}.}
    \label{fig:mass_func_Pr_125}
\end{figure}

We obtained the BFSR to the observational data by using the initial parameters given in Table \ref{tab:Ini_data_best_fit} and the following broken power law IMF, also shown in Fig. \ref{fig:mass_func_Pr_125}:
\begin{equation}
\label{eq:mass_func}
\frac{\xi(m)}{dm}\propto
\begin{cases}
 m^{-0.4}& 0.08 < m/M_\odot < 0.5 \\
 m^{-2.5}& 0.5 \leq m/M_\odot < 1.5 \\
 m^{-4.1}& 1.5 \leq m/M_\odot < 100. \\
\end{cases}
\end{equation}

\begin{table}[h]
\caption{Initial parameters of the BFSR.}
\centering
\begin{tabular}{lll}
\hline
Property & Unit & Value \\
\hline
\hline
Mass  & $\msun$ & 7500  \\
Number of stars & & 15001 \\
Jacobi radius & pc & 27.0  \\
Filling factor $\lambda$ &  & 0.12 \\
Half-mass radius & pc  & 3.24 \\
Plummer radius & pc  & 2.48 \\
Star formation efficiency &  & 17\% \\
\hline
\end{tabular}
\label{tab:Ini_data_best_fit}
\end{table}

\begin{table}[h]
\caption{Overview of the cluster properties in the BFSR at the age of $708 \, \mathrm{Myr}$ in comparison to the observed data of Praesepe.}
\centering
\begin{tabular}{lccc}
\hline
Parameters & Unit & Simulated & Observed \\
\hline
\hline
Jacobi radius $r_J$ & pc & 11.47 & 11.49 \\
filling factor $\lambda$ & & 0.38 & 0.41\\
$M_J(r_{J,\text{sim}})$ \ & $\msun$ & 558 & 566 \\
$N(r_{J,\text{sim}})$ & & 945 & 933 \\
$M_J(r_{J,\text{sim}}),\, m\geq 0.35 \msun$ & $ \msun$ & 487 & 479 \\
$N(r_{J,\text{sim}}),\, m\geq 0.35 \msun$ & & 597 & 605 \\
White dwarfs & & $10$ & $13$ \\
\hline
\end{tabular}
\label{tab:data_best_fit}
\end{table}
For the BFSR, we determined a constant angular momentum of the cluster centre $L_\mathrm{c} = -1883 \,\mathrm{kpc\,km\,s^{-1}}$, the guiding radius $R_\mathrm{g} = 8.0 \, \mathrm{kpc}$, and the angular velocity $\Omega_\mathrm{g}= -30\, \mathrm{rad}\,\mathrm{Gyr}^{-1}$.

The cluster centre is no longer well determined after 1.53 Gyr, which we use as dissolution time. For comparison, we conducted a circular orbit simulation run (COSR) using the same guiding radius and angular momentum as the BFSR. All other initial properties align with the data provided in Table \ref{tab:Ini_data_best_fit}. Some characteristic properties of the BFSR at the current age of Praesepe are given in Table \ref{tab:data_best_fit}. The number of white dwarfs of the ensemble of 15 random realisations with the same initial parameters is $11.1 \pm 3.5$.
 
The Jacobi mass and radius are calculated according to \citet[][see Eq. (A2)]{just2009quantitative} using the normalised epicyclic frequency $\beta=1.37$ and the current angular velocity, $\Omega$, at each time of the cluster simulation. In Fig. \ref{fig:Mj_lambda_over_t}, the Jacobi mass of the cluster (top panel) and the ratio of the half-mass radius to the Jacobi radius, $\lambda$, (bottom panel) are shown as a function of age. The cluster loses most of its mass during violent relaxation in the first 20 Myr as discussed in \citep{shukirgaliyev2017impact, Shukirgaliyev2018,bek+2019, Jerabkova2021}. 
The oscillations of $\lambda$ are mainly due to the variation of the Jacobi radius along the epicyclic motion of the cluster. The evolution of the COSR shows a slightly lower mass loss.

\begin{figure}
    \resizebox{\hsize}{!}{\includegraphics{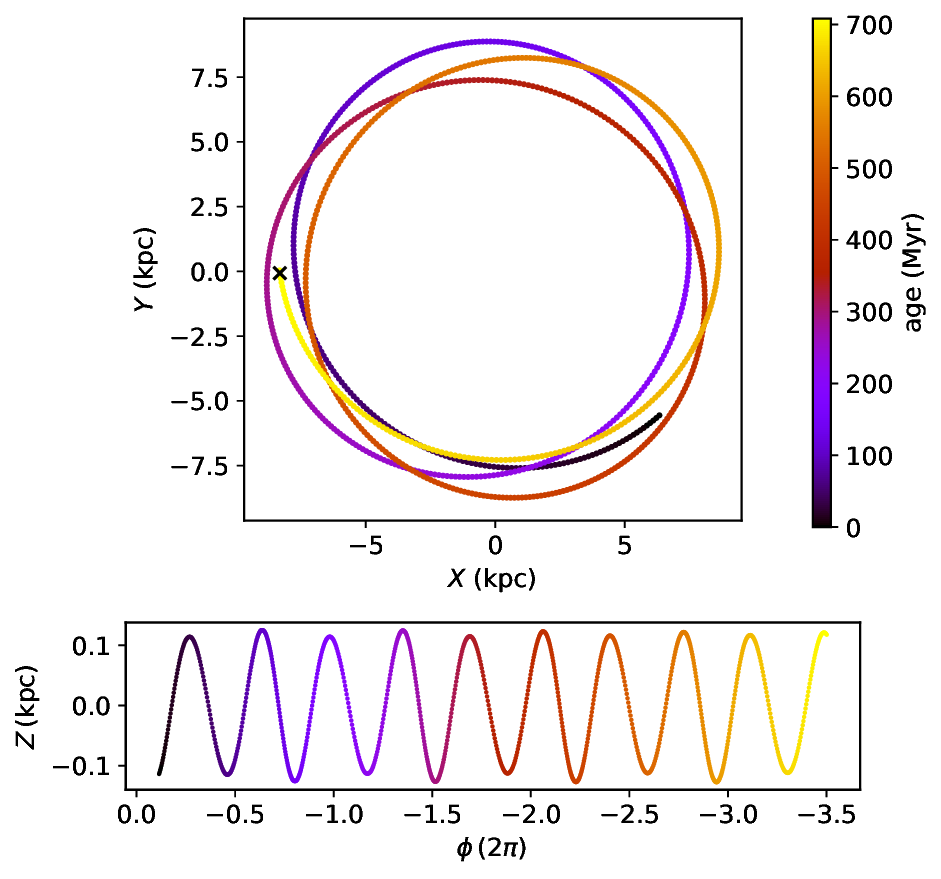}}
    \caption{\textit{Top panel}: Orbit of the simulated cluster in the $X$--$Y$-plane. The black cross marks observed position of today. \textit{Bottom panel}: Motion of the cluster centre in the $Z$--$\phi$-plane.}
    \label{fig:Z_vs_phi}
\end{figure}

\begin{figure}
    \resizebox{\hsize}{!}{\includegraphics{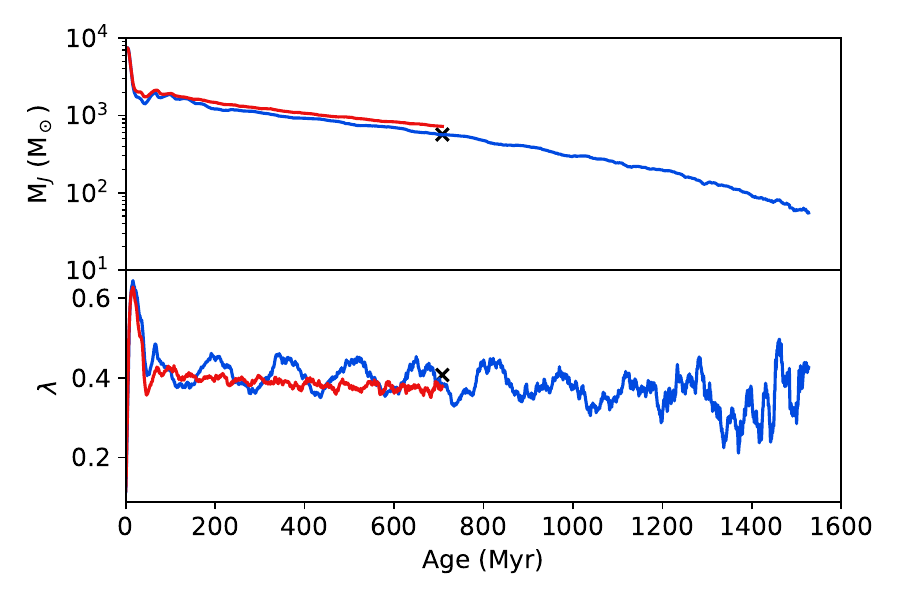}}
    \caption{Evolution of Jacobi mass, M$_J$, and ratio of half-mass radius to Jacobi radius, $\lambda$, as a function of age. \textit{Top panel}: M$_J$ for the BFSR (blue) and the COSR (red). \textit{Bottom panel}: $\lambda$ for the BFSR (blue) and the COSR (red). The observed values at $708\, Myr$ are marked with a black cross. The data of the BFSR are shown until the cluster dissolves after 1.53 Gyr.}
    \label{fig:Mj_lambda_over_t}
\end{figure}

\subsection{Stellar mass function}

The PDMF of the observed cluster and the best simulation are shown in Fig. \ref{fig:mass_func_Pr_125}. Also plotted are the Kroupa IMF \citep{Kroupa2001} and the IMF used for the best simulation, re-scaled for good visibility. The fits to the profiles have breakpoints at $0.5\,\msun$ and $1.5\,\msun$. For the fit to the observed mass function, only bins with centres above the completeness limit are considered. For the fit of the simulation, all bins are considered. Above the completeness limit, model and data fit well. Further, our IMF is consistent to the data down to $0.2\,\msun$. In the BFSR, the total mass of stars inside the Jacobi radius below the completeness limit is $71\,\msun$ and $24 \, \msun$ below $0.2 \, \msun$. For the observation, the respective masses are $68 \, \msun$ and $10 \, \msun$ leading to an agreement of $96\%$ (below $0.35\, \msun$) and $41\%$ (below $0.2\, \msun$). In simulations with more low mass stars (equivalent to a more negative exponent in Eq. (\ref{eq:mass_func}) for the mass range of stars lighter than $0.5\,\msun$), the final profile above the completeness limit does not fit well. Changing the IMF also has an effect on all the other cluster parameters, since the number of stars of different mass affects the cluster dynamics, such that the other initial parameters need to be re-determined.

\subsection{Cumulative mass profile}

\begin{figure}
    \resizebox{\hsize}{!}{\includegraphics{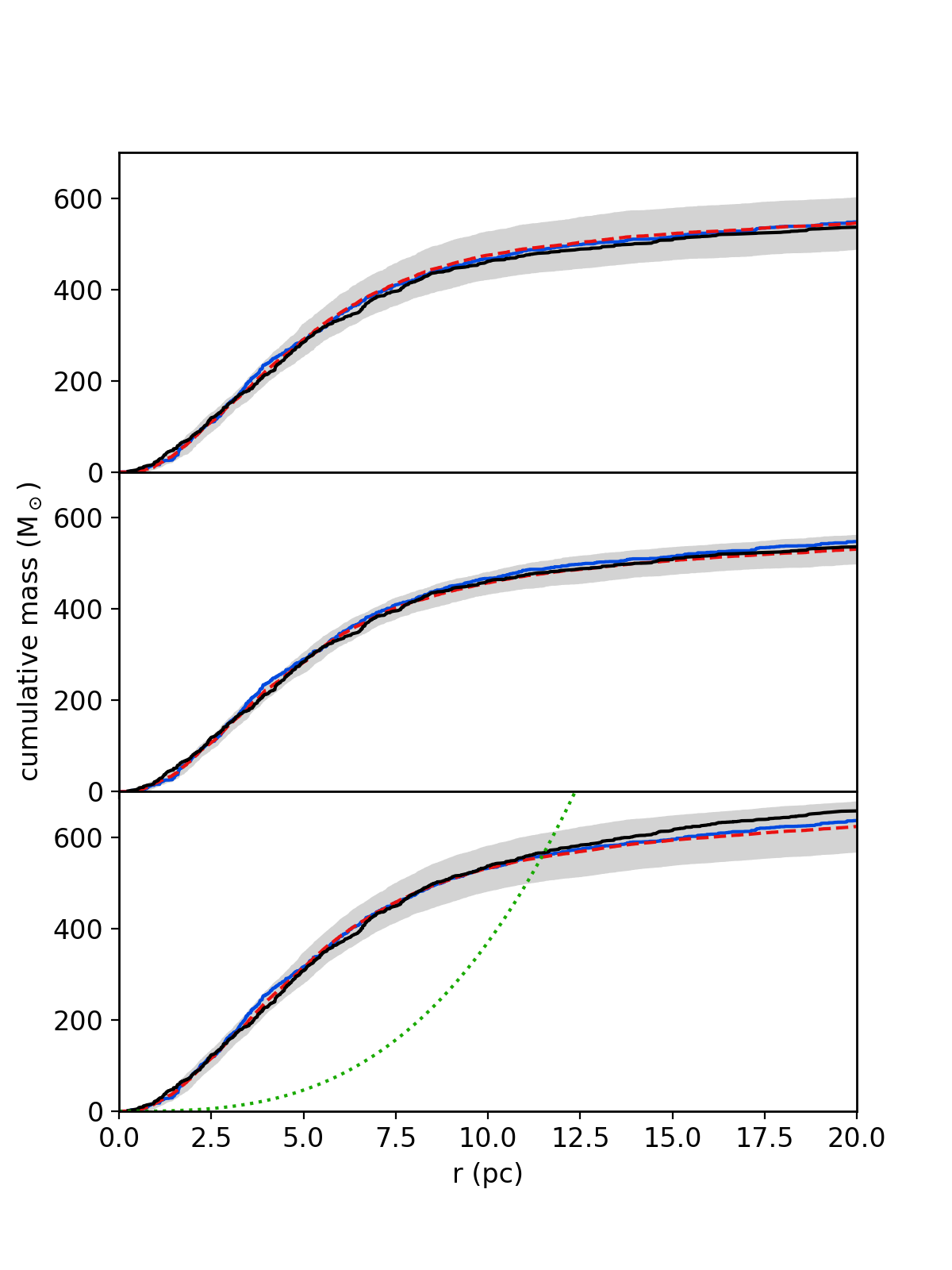}}
    \caption{
    Three panels depicting cumulative mass profiles of stars at {$708 \, \mathrm{Myr}$} under different scenarios: BFSR (solid blue), observational data (solid black), and ensemble averages from multiple simulations with varied parameters (dashed red with $1\,\sigma$ uncertainty in grey). \textit{Top and centre panels}: Stars above the completeness limit derived from a random realisation of the distribution of initial coordinates and velocities, as well as the IMF, respectively. \textit{Bottom panel}: Extended to all stars, averaging all 15 simulations. It also shows the Jacobi mass, M$_\mathrm{J}$(r), as a function of the Jacobi radius (by inverting Eq. (\ref{eq:jacobi_radius}), dotted green).}
    \label{fig:cum_mass_all}
\end{figure}

The upper two panels of Fig. \ref{fig:cum_mass_all} show the cumulative mass profiles of stars above the completeness limit of $0.35 \, \msun$ from the BFSR and seven different runs averaged with different random seeds for the coordinate and velocity distribution (top panel) and the IMF (centre panel), respectively. The dashed red line denotes the mean, and the grey shaded line the $1\,\sigma$ region of the different mass functions. The BFSR and the observational data are also plotted. Compared to the scattering of the profiles due to the different random seeds, the simulations agree very well with the observational profile. 

The bottom panel in Fig. \ref{fig:cum_mass_all} shows the cumulative mass profiles of all stars. In addition to the BFSR and the observational profile, the mean and scatter of all 15 simulations with the same initial conditions but different random seeds are shown. The dotted green line represents the Jacobi mass as a function of the Jacobi radius. The intersection of the curve with the cumulative mass profiles is then at the Jacobi radius of the respective cluster. 

\subsection{Mass segregation}

\begin{figure}
    \resizebox{\hsize}{!}{\includegraphics{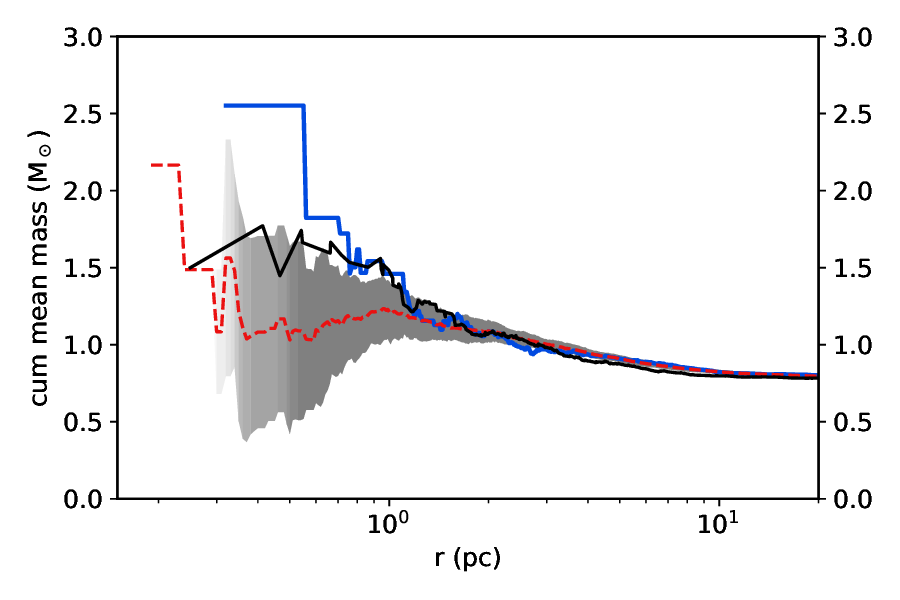}}
    \caption{Mass segregation represented by the cumulative mean mass profile of stars above the completeness limit at {$708 \, \mathrm{Myr}$}: from the BFSR (solid blue), from the observation (solid black), the average of 15 different simulations with different random seeds for the coordinate and velocity distribution or IMF (dashed red), and their $1\,\sigma$ range (grey area). The lighter shades of grey represent the lower number (only $\sim$4 -- 5) of simulations, which had stars inside that radius from the cluster centre.}
    \label{fig:mass_seg_completeness}
\end{figure}

Fig. \ref{fig:mass_seg_completeness} shows the cumulative mean mass inside of a radius as a function of radius for stars above the completeness limit. As in Fig. \ref{fig:cum_mass_all}, the plot includes the profiles for the BFSR, the observational data, and the mean and dispersion of the set of 15 simulations with different random seeds. The mean mass can be interpreted as a measure of mass segregation. For radii larger than $ \sim 1 \, \mathrm{pc}$, the cumulative mean mass profiles of the observations and the BFSR show a clear decreasing trend from $\sim 1.5 \, \msun$ to $0.82 \, \msun$ and are consistent with each other. The innermost $1\,\mathrm{pc}$ is dominated by low number statistics with a flat profile for the observations and with a massive star in the centre for the BFSR. The ensemble average shows a flatter profile in the centre, but the observational data are consistent with the blue line at the $1\,\sigma$ level. That means that the mass segregation can be purely dynamical by 2-body scattering and there is no need for an initial mass segregation.

\section{Tidal tails}\label{sec:results}

\subsection{Structure}\label{sec:structure}

\begin{figure}
    \resizebox{\hsize}{!}{\includegraphics{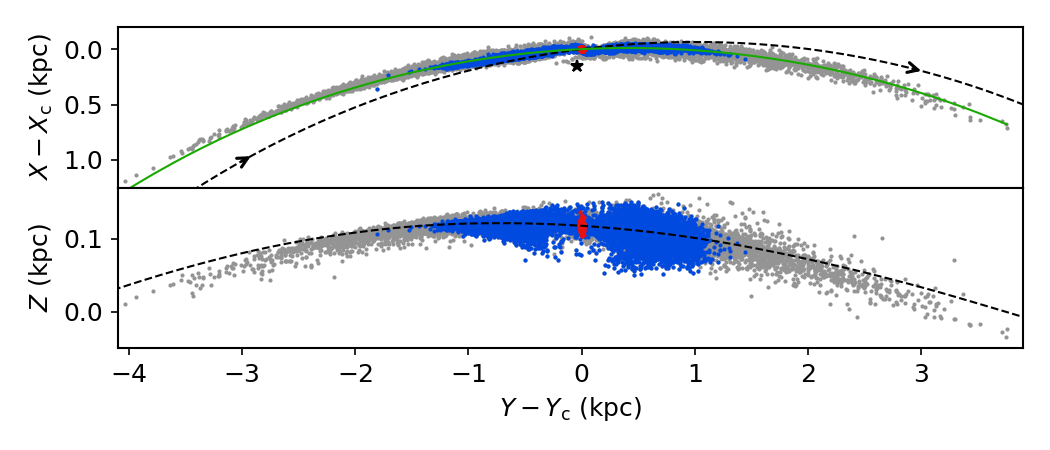}}
    \caption{Structure of the tidal tails at {$708 \, \mathrm{Myr}$}. Blue dots show the stars that escaped the cluster after the violent relaxation phase ended. Grey dots mark all stars that were lost during the violent relaxation phase, lasting the first $20 \, \mathrm{Myr}$ of the simulation. Red dots show the cluster stars and the dashed black line the orbit of the cluster centre with the direction of the orbit indicated. The fitted tail line is shown in green. The Sun is marked with a black star. Note the different scaling of the $Z$-axis. A similar type of tail structure (based on the idea of Tail I -- 'gas expulsion tail' and Tail II -- 'stars evaporation tail') is also described in \cite{Dinnbier2020a}.}
    \label{fig:structure_tails}
\end{figure}

The structure of the tidal tails can be seen in Fig. \ref{fig:structure_tails}. At $708 \, \mathrm{Myr}$, the cluster is located between the apo- and pericentre moving towards the pericentre. The arms containing all stars (blue and grey dots) extend about $4 \, \mathrm{kpc}$. The tails of the stars that have escaped after the violent relaxation phase (blue dots) extend about $1.2\,\mathrm{kpc}$. We can identify the leading (positive $Y-Y_\mathrm{c}$) and trailing (negative $Y-Y_\mathrm{c}$) tail. Also, the tail line (green), determined in the local cylindrical system in Fig. \ref{fig:average_tail} (top panel) can be seen transformed to the Cartesian system. In vertical direction, seen in the lower panel of Fig. \ref{fig:structure_tails}, the cluster recently has passed the highest deflection point of $120 \, \mathrm{pc}$. We observe an asymmetry in the $Y$--$Z$-plane. For further analysis of the tail structure, we used the stars that escaped the cluster after the violent relaxation phase, coloured blue in Fig. \ref{fig:structure_tails}.

\begin{figure}
    \resizebox{\hsize}{!}{\includegraphics{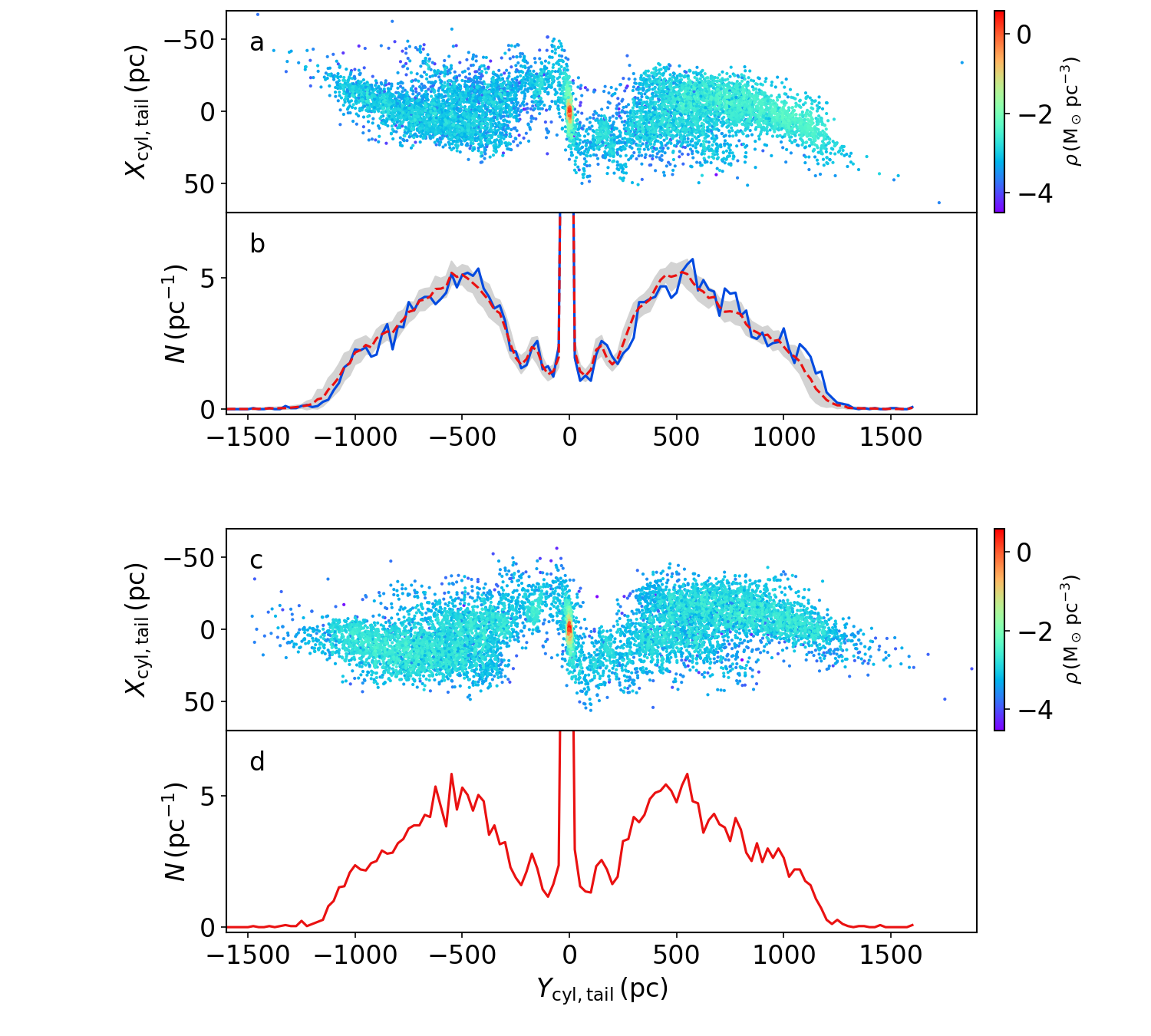}}
    \caption{3D densities and 1D number densities of the tidal tails of the BFSR and the COSR at {$708 \, \mathrm{Myr}$}, with the $x$-axis being the tail line shown in Fig. \ref{fig:average_tail} (top panel). \textit{Upper two panels}: Densities at $708\,\mathrm{Myr}$ of the BFSR. \textit{Lower two panels}: Densities of the COSR at $708\,\mathrm{Myr}$. 
(a) Scatter plot of the tidal tails (blue dots in Fig. \ref{fig:structure_tails}) with colour-coded logarithm of the 3D mass density. (b) 1D Number density of the BFSR (blue line). The dashed red line shows the mean of all 15 simulations. The grey area shows the $1\,\sigma$ region. The panels
    (c) and (d) show the same as for the COSR. Note that in (d) only the 1D number density is shown, as only one run was done for comparison.}
    \label{fig:1D_density_profile}
\end{figure}

\begin{figure}
    \resizebox{\hsize}{!}{\includegraphics{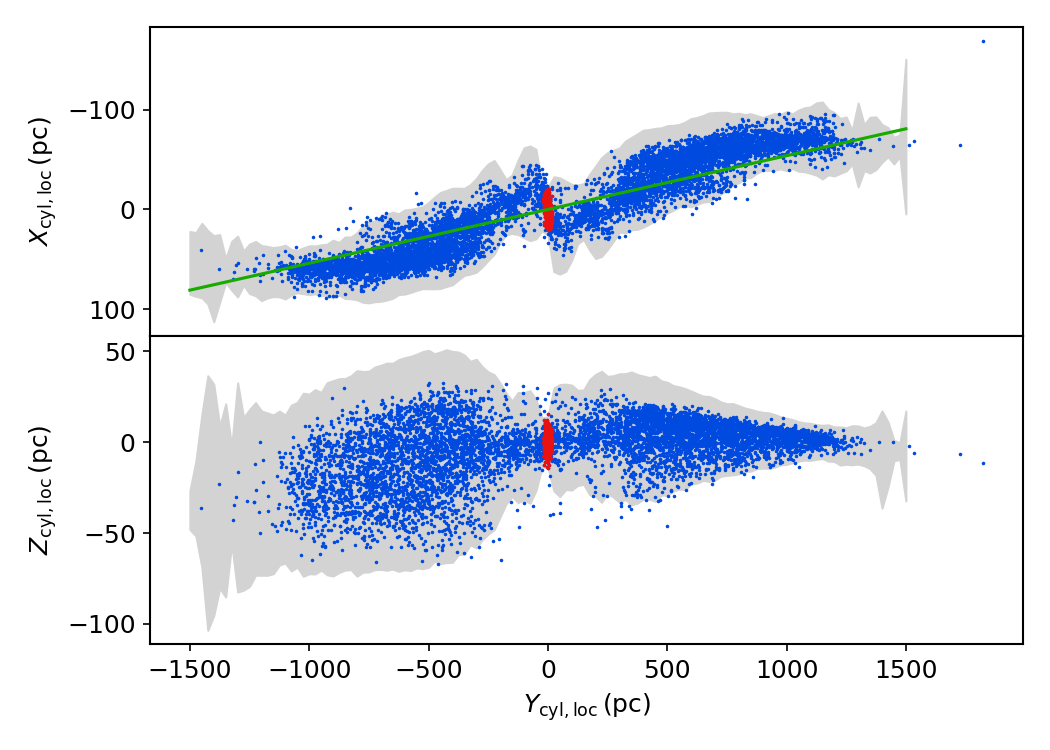}}
    \caption{Average tail shapes at {$708 \, \mathrm{Myr}$}. \textit{Top panel}: $X_\mathrm{cyl,loc}$--$Y_\mathrm{cyl,loc}$-plane shape with the tail line in green. \textit{Bottom panel}: $Y_\mathrm{cyl,loc}$--$Z_\mathrm{cyl,loc}$-plane shape via the grey area, denoting the $3\,\sigma$ region of the stars to the mean per bin. The blue and red dots represent the BFSR tail stars and the cluster stars, respectively. Note the scaling of the ordinate axis in each of the panels.}
    \label{fig:average_tail}
\end{figure}

A density analysis is shown in Fig. \ref{fig:1D_density_profile}. We observe an asymmetry in the tails' 3D density (top panels) but not in the corresponding tail in the 1D number density along the tail line. It can be explained with the observed asymmetry of the vertical width of the tails, in Fig. \ref{fig:structure_tails} (lower panel). The trailing tail is more compact regarding the vertical spread, while the leading tail is more diffuse. This, of course, leads on average to a higher density in the trailing arm as can be seen in the right arm in Fig. \ref{fig:1D_density_profile} (a). In the Galactic plane, we do not see this asymmetry. For comparison, the resulting 3D mass density and 1D number density of the COSR is shown in the lower two panels in Fig. \ref{fig:1D_density_profile}. There, no 3D mass density asymmetry can be seen, which is expected, since the cluster on a circular orbit does not have oscillation in vertical direction. We observe a similar distribution in the 1D number density as for the eccentric orbit, compare Fig. \ref{fig:1D_density_profile} (b) and (d). We can distinguish the first clump at about $\pm 160 \, \mathrm{pc}$ and a broad maximum with the peak around $500\,\mathrm{pc}$. This is different from the finding of \citet{just2009quantitative,kupper2008structure} for a circular orbit with a nearly constant mass loss rate and more visible clumps in the tails. In \citet{kupper2010tidal} broad maxima were also found in the tail density profiles for an eccentric orbit and also a few clumps in each tail could be identified. A possible reason could be the lower mass of our cluster, which leads to a larger noise in escape velocities. This corresponds to different phases in the stars' epicyclic motion, which could smear out the clumps or even prevent their formation.

In Fig. \ref{fig:1D_density_profile} (b), the statistical evaluation of different initialisation seeds for the simulations, on one hand, indicates that the 1D number density is robust with respect to the choice of the seed and, on the other hand, that the BFSR is in very good agreement with the mean over all 15 simulations. For comparison, the result of the COSR is shown Fig. \ref{fig:1D_density_profile} (d).

An analysis of the average tail shape is presented in  Fig. \ref{fig:average_tail}. Mean and standard deviation (grey area) of the stars radial and vertical coordinate are calculated per bin using a bin width of $25\,\mathrm{pc}$. We conclude from this that the tail shape in the radial and vertical plane is not sensitive to the choice of initialisation seed. This is in good agreement with the robustness of the 1D number density.

\subsection{Angular momentum}\label{sec:angmom}

The \textit{z}-component of the angular momentum of a star leaving the cluster is expected to be conserved if the star is far enough away from the cluster. Hence, the \textit{z}-component of the angular momentum should become constant if the star surpasses a certain distance from the cluster. In our simulations, we find that stars far out in the tail still have a significant alteration of the $z$-component of their angular momentum. We quantified the origin of this contribution to $\Delta L$ (Eq. (\ref{eq:DeltaL})) at the example of a typical star (index 3990) in the trailing tail. We compared two methods to calculate the evolution of $\Delta L$. The two should agree with each other as a sanity check. The first method is the direct measurement via the cross product of position and velocity of the typical star. In the second method, we calculated at each timestep the forces and thus the torques of each star born in the cluster on the typical star. For the analysis of the origin of the resulting change in angular momentum, we built groups of stars as described below.

\begin{figure}
\resizebox{\hsize}{!}{\includegraphics[width=0.9\linewidth]{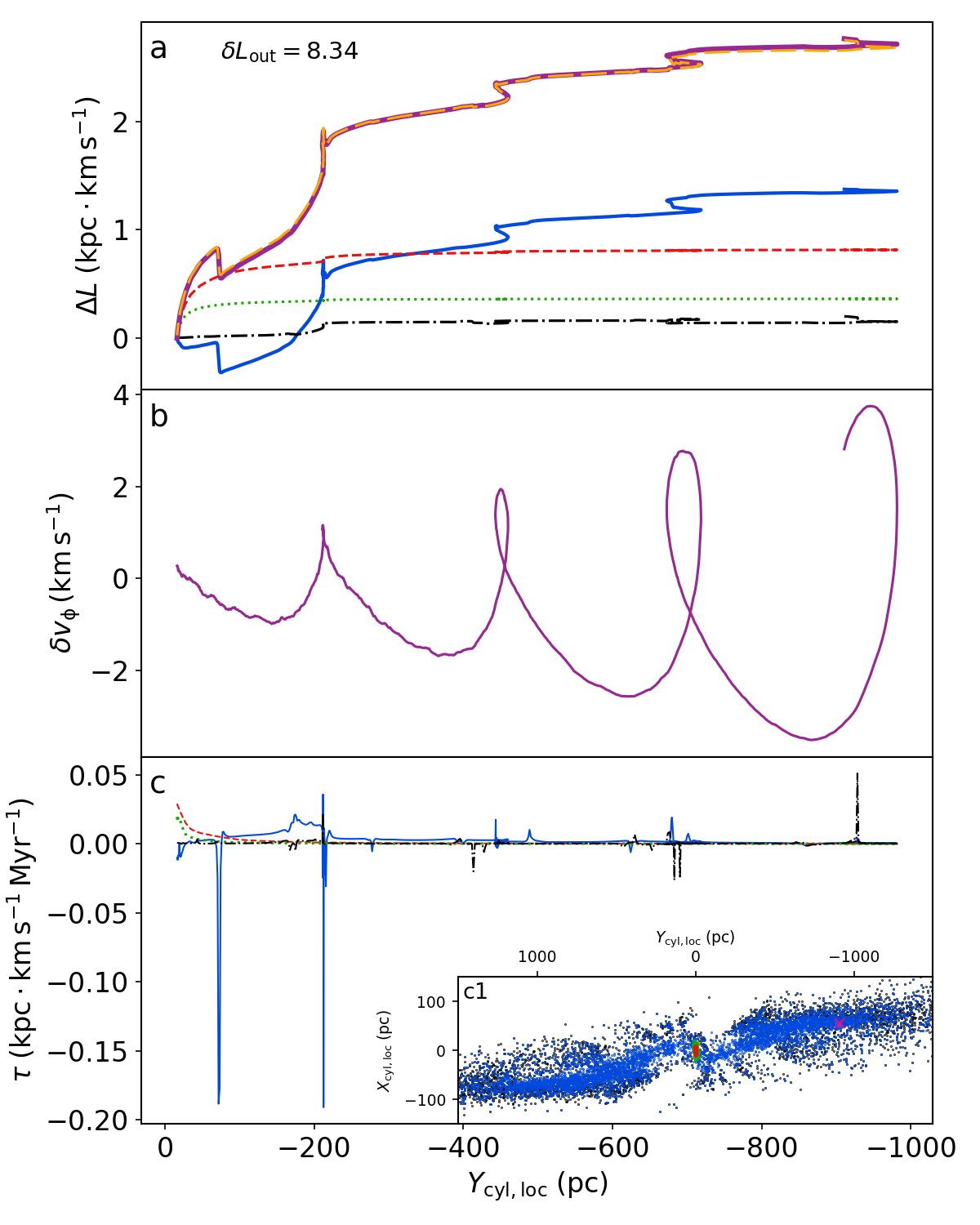}}
   \caption{Properties of a typical star in the tidal tails. The direct measurement of $\Delta L$ of a typical star and the contributions to that $\Delta L$, the tangential velocity $\delta v_\phi$ relative to the cluster centre, and the torques $\tau$ exerted onto the star by different subsystems are all plotted as a function of the 
    distance to the cluster centre, $Y_\mathrm{cyl,loc}$. The different colours in panels (a) and (c), paired with their respective line styles represent the contributions of the different subsystems: by stars inside the cluster (red dashed), by stars in between one and two Jacobi radii (dotted green), by stars lost during the violent relaxation phase (dot-dashed black), and by stars in the tidal tails (blue).
    (a) $\Delta L$ as a result of the time integral of the torques in (c) shown in the corresponding colours and line styles. The direct measurement of $\Delta L$ of the typical star is also indicated with the thick purple line, as well as the sum of all subsystems' $\Delta L$ with the long-dashed amber line. Note that the long-dashed amber line superimposes the thick purple line. The offset in the $z$-component of the angular momentum,  $\delta L_\mathrm{out}$, with respect to the cluster centre at time of escape corresponds to the zero point of $\Delta L$. It is given in  $\mathrm{kpc \cdot km\,s^{-1}}$.
    (b) Tangential velocity relative to the cluster centre of our typical star.
    (c) $z$-component of the torques, $\tau$, of each of the subsystems, exerted onto our typical star. 
    (c1) Inset of the cluster and its tails in the local cylindrical system at {$708 \, \mathrm{Myr}$}. Colouring is according to the description above. Additionally, the typical star is marked with a pink cross.}
    \label{fig:torque_dL_vtang}
\end{figure}

The direct measurement of $\Delta L$ of our typical star is shown with the thick purple line in Fig. \ref{fig:torque_dL_vtang} (a) after leaving the cluster at $t_\mathrm{out}=70\,\mathrm{Myr}$ as a function of distance to the cluster. We can observe that the $z$-component of the angular momentum of this star does not become constant, even at a distance as large as nearly $1 \, \mathrm{kpc}$ to the cluster. Along the way farther out in the tail, the star experiences steps in $\Delta L$ around $-200 \, \mathrm{pc}$, $-450\,\mathrm{pc}$, and $-700\,\mathrm{pc}$, as well as the start of a step at about $-950\,\mathrm{pc}$, where $\Delta L$ is increased rapidly. In time, these steps occur every $157\,\mathrm{Myr}$, which is the epicyclic period of the cluster relative to the Galactic centre and also of the tail stars relative to the cluster centre.
At the time of escape, the star has a $\delta L_\mathrm{out} = 8.34 \, \mathrm{kpc \, km \, s^{-1}}$ (Eq. (\ref{eq:dL_out})) and at the end of the simulation the $z$-component of the angular momentum altered by $\Delta L = 2.76 \, \mathrm{kpc \, km \, s^{-1}}$, meaning that the $z$-component of the angular momentum of the star relative to the cluster centre then is $11.10 \, \mathrm{kpc \, km \, s^{-1}}$. The alteration by $\Delta L = 2.76 \, \mathrm{kpc \, km \, s^{-1}}$ corresponds to an outward shift of $12.5 \, \mathrm{pc}$ in its guiding radius.

The tangential velocity relative to the cluster centre,
\begin{equation}
    \delta v_\phi = v_{\phi,*} - v_{\phi,\mathrm{c}},
\end{equation}
with $v_{\phi,*}$ and $v_{\phi,\mathrm{c}}$ being the co-moving Galactocentric tangential velocity of the star and the cluster centre, respectively, is presented in Fig. \ref{fig:torque_dL_vtang} (b). It shows the difference of the epicyclic motion of the star to the cluster centre's epicyclic motion. It changes sign from negative (movement away from the cluster) to positive (movement towards the cluster) during the angular momentum steps and forms loops. These loops form when the stars are closest to the cluster orbit, which occurs during the pericentre passage on the trailing arm. The maxima of these loops are, such as the steps in $\Delta L$ also occurring every $157\,\mathrm{Myr}$.

These loops show the process of the stars moving closer to the cluster and thus the tidal tails getting compressed temporarily \citep{kupper2010tidal}. Therefore, as the stars are slowed down during the compression, the likelihood of possible slow encounters with nearby stars in the tails is increased, subsequently leading to a significantly larger alteration in the $z$-component of the stars' angular momentum. Additionally, with the loop behaviour of the velocity, we observe that the epicycles around the guiding radius increase in amplitude with an increasing  $\Delta L$ (Fig. \ref{fig:torque_dL_vtang} (a)) leading to a larger thickness of the tails. Combining both effects results in a total radial offset of the guiding radius of up to more than $50 \, \mathrm{pc}$ for a fraction of stars. For a typical star, with a $\Delta L \approx 1.6\, \mathrm{kpc\,km \, s^{-1}}$, this offset of the guiding radius is of around $7\, \mathrm{pc}$.

Fig. \ref{fig:torque_dL_vtang} (c) shows the torques exerted onto our typical star by stars of different subsystems. The different subsystems are: Stars in the cluster centre (dashed red line), stars in between one and two Jacobi radii (dotted green line), stars in the tidal tails (blue line), and stars lost during the violent relaxation phase (dot-dashed black line). The torques are plotted as a function of distance to the cluster centre. For an overview of these subsystems, Fig. \ref{fig:torque_dL_vtang} (c1) shows an inset of the cluster and the tidal tails in the local cylindrical system (Eq. (\ref{eq:loc_cyl})) at $708\,\mathrm{Myr}$ with the colours corresponding to each subsystem as listed before. Additionally, the typical star is marked with a pink cross. In Fig. \ref{fig:torque_dL_vtang} (c), we see that all stars within $2r_\mathrm{J}$ (dashed red and dotted green lines) exert a significant torque on our typical star until it reaches about $200 \,\mathrm{pc}$ from the cluster. The figure further shows that the cluster's torque vanishes, as expected from the potential of the cluster and energy conservation, which in turn means that the influence on the star's $\Delta L$ becomes constant. Only stars in the tail and stars lost in the violent relaxation phase exert a significant torque onto our typical star further away from the cluster. Dominant torque spike groupings, including both smaller spikes with absolute values up to $\sim0.025 \, \mathrm{kpc\,km\,s^{-1}\,Myr^{-1}}$ and larger spikes, arise around $-200 \, \mathrm{pc}$, $-450\,\mathrm{pc}$, $-700\,\mathrm{pc}$, and $-950\,\mathrm{pc}$, which are the same locations of the steps in Fig. \ref{fig:torque_dL_vtang} (a). These spikes could be a net effect of a few slow encounters of our typical star with other stars in the corresponding subsystem, since at these positions the epicyclic motion has its maxima, where clumps usually form \citep{just2009quantitative} and slow down stars, as seen in Fig. \ref{fig:torque_dL_vtang} (b).

Taking the cumulative sum of these torques multiplied with the time step $\delta t = 0.673 \, \mathrm{Myr}$ at each snapshot (Eq. (\ref{eq:torque})), we obtain the $\Delta L$ fraction of the typical star produced by each of the subsystems. These are shown in Fig. \ref{fig:torque_dL_vtang} (a) in addition to the already discussed direct measurement and have the same colours and line styles as in Fig. \ref{fig:torque_dL_vtang} (c). The sum of all the contributions of the subsystems is shown with the long-dashed amber line.
The cluster stars' and in-between cluster and tails stars' contribution to $\Delta L$ becomes constant after about $200 \, \mathrm{pc}$ distance to the cluster, which is consistent with Fig. \ref{fig:torque_dL_vtang} (c). Stars lost during the violent relaxation phase (dot-dashed black line) have a diminutive but non-negligible influence on $\Delta L$. The dominant contribution, leading to a growing $\Delta L$, is given by stars in the tidal tails (blue line) showing that the steps are produced predominantly  by these stars.
The direct measurement of $\Delta L$ (thick purple line) and the sum of all contributions of the subsystems (long-dashed amber line) in Fig. \ref{fig:torque_dL_vtang} (a) coincide very well. This shows that far out in the tail, the $z$-component of the angular momentum of our typical star is significantly altered by the torques it experiences with other tail stars and stars lost during the violent relaxation phase during its travel away from the cluster. The tail stars contribute around $36\,\%$ to the $\Delta L$ of our typical star, while the tail stars' contribution generally, in the BFSR, can be up to $70\,\%$. The slow encounters, seen in the torque spike groupings in Fig. \ref{fig:torque_dL_vtang} (c), form the steps in $\Delta L$ around $-200 \, \mathrm{pc}$, $-450\,\mathrm{pc}$, and $-700\,\mathrm{pc}$, and $-950\,\mathrm{pc}$, as seen in Fig. \ref{fig:torque_dL_vtang} (a). Additionally, we note that the direction of the torque onto this star, in total, remains the same during its travel out in the tail. This behaviour in $\Delta L$ can be observed for all tidal tail stars of the BFSR in the leading and trailing tail. A random selection can be seen in the left column of Fig. \ref{fig:appendix_angmom_plot}.

This analysis shows that the self-gravity of the tidal tails on an eccentric orbit, despite its very low volume density, has a non-negligible effect on the dynamics of the tails' stars and consequently on the shape and structure of the tails. This, however, is not an effect of an eccentric cluster orbit. Tail stars on a circular orbit exhibit a similar behaviour, as can be seen in the right column of Fig. \ref{fig:appendix_angmom_plot}, which shows the $\Delta L$-behaviour of stars from the COSR. In general, on both orbits, the $\Delta L$ of the tail and the cluster stars can have different signs, i.e. the cluster stars can have a positive $\Delta L$ and the tail stars a negative $\Delta L$ effect. This may have a smearing effect in the density distribution or the contrast between clump and no clump, such that we cannot see them in the density distributions.

Summarising, we cannot identify clumps in the density, but we observe indications in position and velocity. Based on the position of the steps, we can determine the distance between each step, denoted as $d_\mathrm{step}$, and compare it to the predicted distance $y_\mathrm{T}$ between clumps using Eq. (22) in \cite{just2009quantitative}. The comparison is presented in Table \ref{tab:comparison_distance_clumps} by inserting the $z$-component of the angular momentum of the star $L_*$ relative to the cluster centre $L_c$, $\delta L = L_* - L_c = \delta L_{\rm out} + \Delta L$, during the step into the equation for $y_\mathrm{T}$. The predicted distances are in agreement with the measured distances from the simulation. With $y_\mathrm{T}$, we also predict the next step, which starts at around $-950\,\mathrm{pc}$ (see Fig. \ref{fig:torque_dL_vtang} (a)).

\begin{table}[h]
\caption{Comparing distances of clumps via direct measurement and theoretical calculation.}
\centering
\begin{tabular}{cccc}
\hline
Step (pc) & $\delta L \, \mathrm{(kpc\,km\,s^{-1})}$ & $d_\mathrm{step} \, \mathrm{(pc)}$ & $y_\mathrm{T} \, \mathrm{(pc)}$ \\
\hline
\hline
-211 & 10.58 & 239 & 247 \\
-450 & 10.88 & 250 & 254 \\
-700 & 11.06 & --- & 259 \\
\hline
\end{tabular}
\label{tab:comparison_distance_clumps}
\tablefoot{
 The distance to the next step from the one at $-700\,\mathrm{pc}$ was not estimated, as at the end of the simulation the star had yet not passed through the entire step.
}
\end{table}

\begin{figure}
    \resizebox{\hsize}{!}{\includegraphics{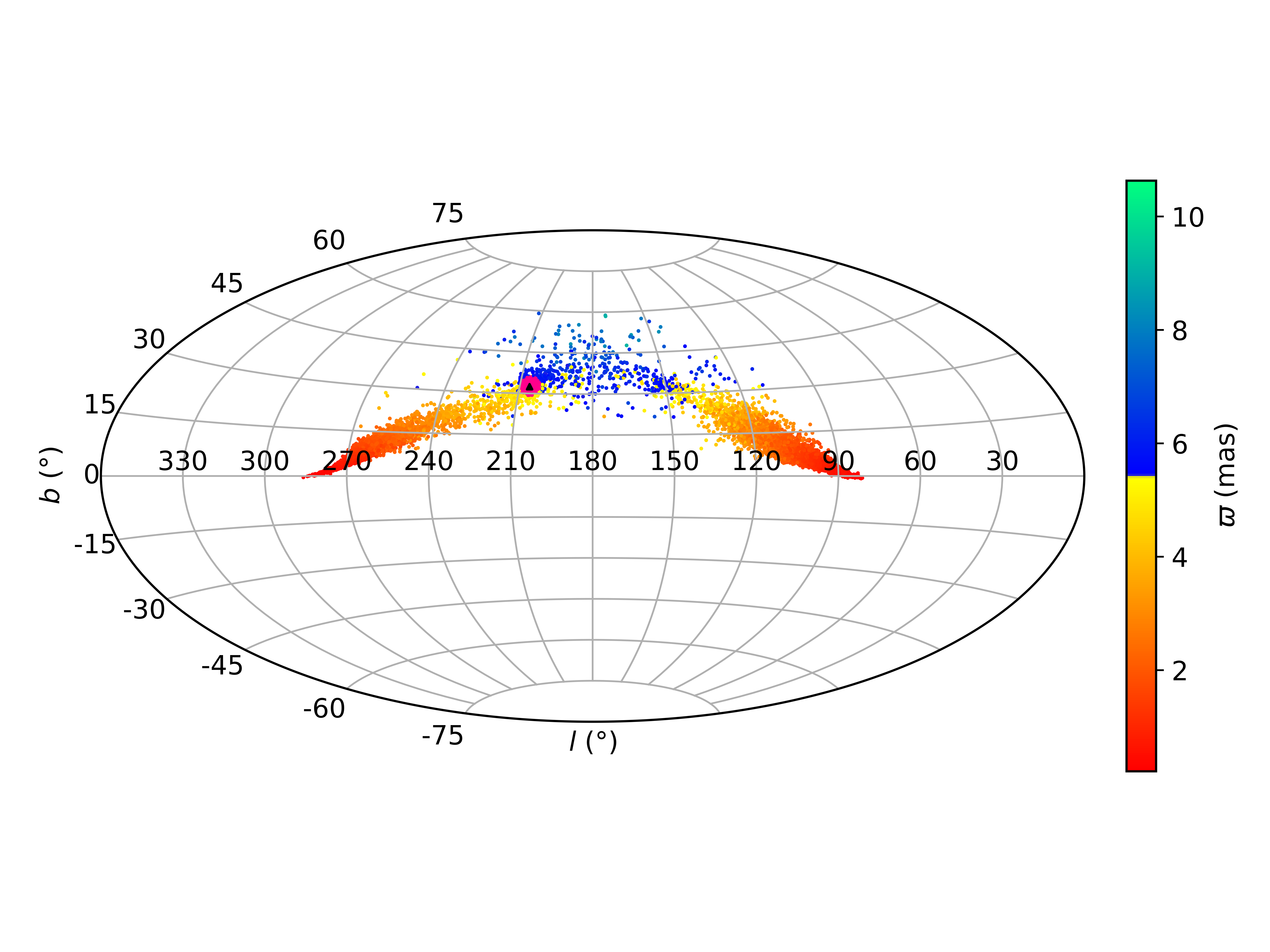}}
    \caption{Aitoff projection of the cluster and its tidal tails across the sky with colour-coded parallax in Galactic coordinates at {$708 \, \mathrm{Myr}$}. The leading arm can be seen on the right and the trailing arm on the left. The split on the colour bar lies exactly at the parallax of the cluster centre at $\varpi = 5.424 \, \mathrm{mas}$. The cluster stars are coloured as pink dots. The observed cluster centre lies at $(l,b)=(205.9\degr,32.5\degr)$ and is marked with a black triangle.}
    \label{fig:aitoff}
\end{figure}

\subsection{Prediction of the tails in the sky}\label{sec:prediciton}

A search for tidal tail stars of Praesepe has already been performed by \citet{roser2019praesepe}. They identify the leading arm up to a cluster distance of about 150 pc and find only a very weak signal for the trailing arm due to the smaller parallaxes of these stars. With the improved proper motions and parallaxes of Gaia (E)DR3 compared to DR2, the search for tail members may be extended to a distance of 250 pc from the Sun.

In \citet{roser2019praesepe}, the data sample was compared to a $N$-body simulation on a circular orbit simply shifted to the cluster position. Here, we present a tailored simulation of Praesepe on an eccentric orbit including the epicyclic motion in the Galactic plane and the vertical oscillations. Therefore, our predictions are much more precise for a future search of tidal tail members.
Using the latest snapshot at $708 \, \mathrm{Myr}$ from the BFSR, we made predictions about the potential position and distance of the tidal tails of the Praesepe cluster on the sky, as shown in Fig. \ref{fig:aitoff}. Here we shift the cluster centre of the simulation to the observed one (see Sec. \ref{sec:obs_data}). All other stars are shifted according to the shift of the cluster centre, before transforming to Galactic coordinates. Fig. \ref{fig:aitoff} shows that the tidal tails span across the entire sky. 

The trailing arm (left arm) was too far away to be observed in Gaia DR2 data. But the leading tail (right arm) is closer to the Sun, as can be seen in Fig. \ref{fig:aitoff} with the transition from blue to orange in the parallax. Comparing Fig. 5 in \cite{roser2019praesepe} and the blue coloured stars in Fig. \ref{fig:aitoff}, we see a good match. The leading tail stars close to the Sun near the cluster, stars that have recently escaped, confirm the stars found in \cite{roser2019praesepe}.
Overall, the parallax for the tidal tails ranges from 1 to $7\, \mathrm{mas}$. This range is measurable by Gaia, subject to the parallax error restrictions. Note that the orange and red stars of both arms cannot be found because they are lying in the Galactic plane. There, the density of field stars is too high, to be able to separate these stars.

\begin{figure}
    \resizebox{\hsize}{!}{\includegraphics{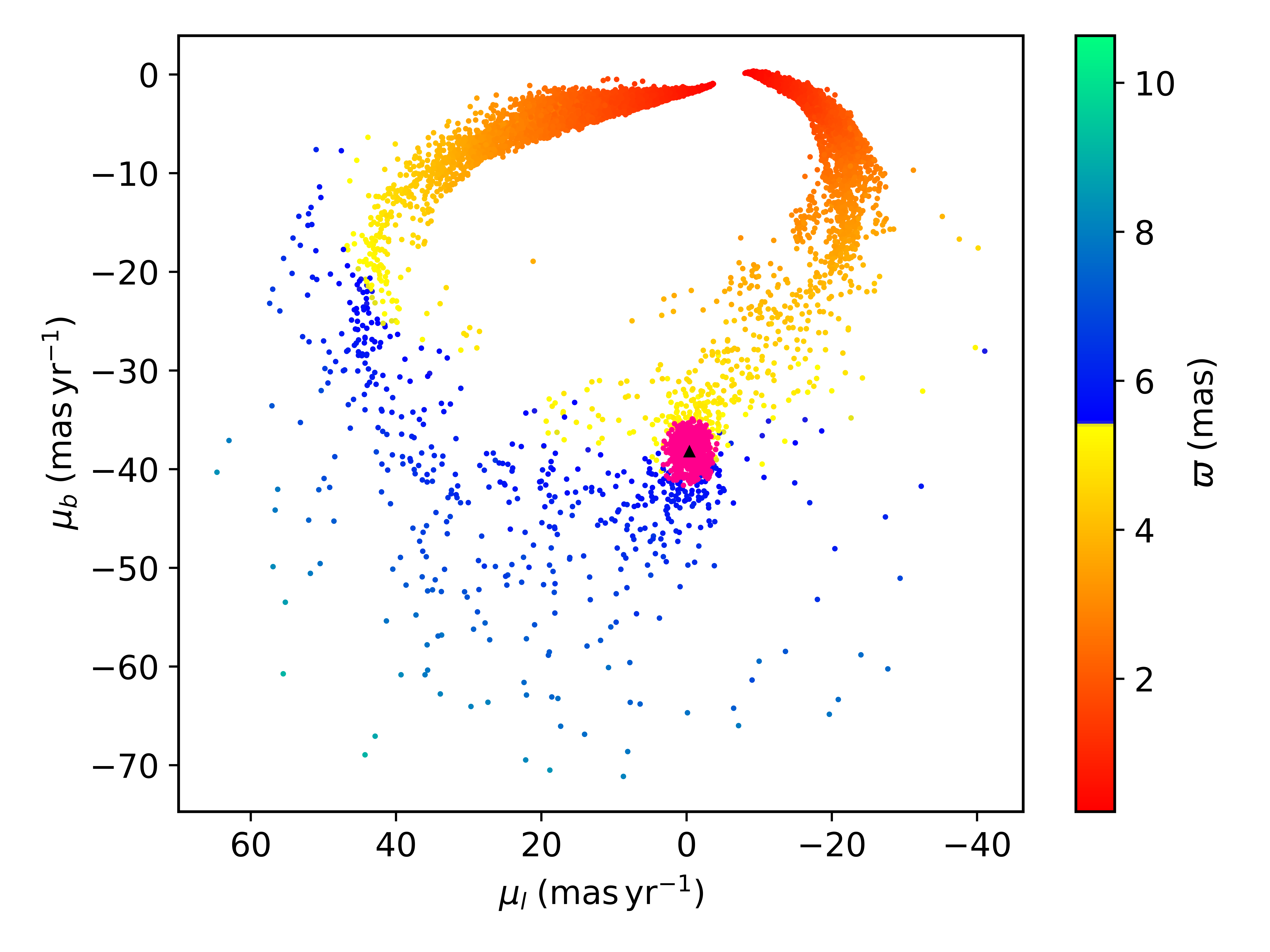}}
    \caption{Distribution of the stars in the proper motion plane with colour-coded parallax at {$708 \, \mathrm{Myr}$}. The left and the right arm are the leading and trailing arm, respectively. The split on the colour bar lies exactly at the parallax of the cluster centre at $\varpi = 5.424 \, \mathrm{mas}$. The cluster stars are indicated with pink dots. The proper motion of the observed cluster centre is marked with a black triangle.}
    \label{fig:cluster_lb_mub}
\end{figure}

Additionally, we present the proper motions in Fig. \ref{fig:cluster_lb_mub}. We shifted the positions and velocities onto the observed cluster. After correction for the Sun's movement, we transformed to the proper motions. In the proper motion plane, both arms and the cluster are distinguishable. These properties help us constrain the stars that can be considered as part of the tidal tails. The yellow coloured stars, which are shifted significantly in $\mu_b$ relative to the cluster, may be observable in Gaia DR3 data.

\section{Summary and conclusions}\label{sec:conc}

Our study introduced a model with a defined set of initial parameters tailored to reproduce the different cluster properties observed in Praesepe. Using direct numerical simulations, we used the \PGPU code to simulate the dynamical evolution of the cluster in order to derive estimations for its initial parameters. Observational data from Gaia EDR3 were used to compare and evaluate how well the different simulations match the observed reality. We compared cumulative mass profiles, the mass segregation, and the stellar mass function using stars above the observational completeness limit ($0.35 \, \msun$). 

Our best model, BFSR, has an initial mass of $7500\, \msun$, a global SFE$_\mathrm{gl}=17\%$ taken from \cite{shukirgaliyev2017impact, Shukirgaliyev2018, bek+2019}, an IMF with boundary masses at $0.08~\mathrm{M_\odot}$ and $100~\mathrm{M_\odot}$, and break masses at $m_1=0.5~\mathrm{M_\odot}$ and $m_2=1.5~\mathrm{M_\odot}$ using the exponents $\alpha_i=[0.4,~2.5,~4.1]$ and an initial filling factor of $\lambda=0.12$. The cumulative mass profile including the cluster mass and the half-mass radius are very well reproduced by the BFSR out to twice the Jacobi radius. With the adapted broken power law IMF, the PDMF including the number of white dwarfs is well reproduced. Compared to the Kroupa IMF, the new IMF has steeper profiles at the low- and high-mass ranges, reducing the contribution to the total mass at both ends. The upper break point is at $1.5\,\msun$ instead of $1\,\msun$. The present-day mass segregation depends strongly on the random distribution for positions, velocities and masses of the stars. The observed mass segregation is at the $1\,\sigma$ limit of the distribution. The BFSR shows an even higher mass segregation in the inner $1\,\mathrm{pc}$ of the cluster. Therefore, no initial mass segregation is required.

One major simplification in the simulations presented here are the missing primordial binary stars. However, for the case of the Hyades with similar mass and age as Praesepe, \citet{ernst+2011} demonstrated that the main effect of including 33\% binaries was a moderate reduction of the initial cluster mass by 5\%. On the other hand, the results on the IMF and mass segregation should be taken with caution. Adding a large fraction of binaries to the initial conditions may alter the best-fit IMF and the degree of mass segregation (see also \cite{Wirth2024}.

To assess the robustness of our model, we ran 14 additional simulations -- seven initiated with different random configurations of initial positions and velocities, and seven different realisations of initial stellar mass. Among these simulations, our most accurate representation closely matches the average behaviour observed across the ensemble of randomly initialised instances. We observe that the overall shape of the tidal tails is not sensitive to the choice of the initialisation seeds for the mass and coordinate distribution. In vertical direction, the tidal tails bend along the orbit taken by the cluster centre, leading to asymmetries in that direction. However, such asymmetries are not evident when projected onto the Galactic plane, indicating no sensitivity to the 1D number density.

Only two tail clumps are identified near the cluster of the BFSR, contrary to the findings of \cite{just2009quantitative}, \cite{kupper2008structure}, \cite{kupper2010tidal} and \cite{kupper2012more}, which showed the formation of multiple distinguishable clumps on both circular and eccentric orbits. Farther out, the clumps found are superposed by a broad maximum in the 1D number density, which may result from the different escape velocities the stars have, leading to different phases in their epicyclic motion. This phenomenon may either smear out the clumps or prevent their formation. Additionally, with the BFSR, we qualitatively matched the area of the identified tail stars in \cite{roser2019praesepe}. These consist of the stars closest to the Sun on the passing leading arm. At larger distances, the leading and trailing arms are harder to identify due to the increasing impact of observational errors and the closer vicinity to the Galactic plane.

We conducted a comprehensive analysis of the structure and dynamics of the tidal tails in our model. A key finding is the self-gravitational effect, which the tidal tail stars exert on themselves, leading to a significant alteration in the \textit{z}-component of the angular momentum -- constituting up to $70\%$ of the overall change after leaving the cluster. Minor contributions come from cluster stars, stars located between the cluster and tails, and those lost during the violent relaxation phase. This contradicts the intuitive assumption that the \textit{z}-component of the angular momentum of a star is constant after it escapes.

The most substantial changes occur during the slow motion phase of the cycloidal orbit of the stars in the system co-rotating with the cluster. Notably, this behaviour is not exclusive to an eccentric orbit; it is also observed in the case of a circular orbit.
The self-gravitating effect of the tail stars leads to a systematic shift in tidal tail streamline centre position, velocity, and width. If not taken into account, it may disturb the search for tail stars in the phase space of Gaia DR3 data. The width or scatter in velocity of the tidal tail stars can be used to estimate the velocity dispersion and mass of the corresponding cluster, which may then be biased.

The stellar density of field stars in the solar neighbourhood is $\sim 0.05 \msun\,\mathrm{pc}^{-2}$, which is a factor of a few higher than the density of tail stars in the clumps (see Fig. \ref{fig:1D_density_profile}). But the efficiency of orbit diffusion in phase space is much less efficient, because of the more symmetric (homogeneous) density distribution and the much larger encounter velocities with tail stars.

\begin{acknowledgements}

The authors thank the anonymous referee for a very constructive report and suggestions that helped significantly improve the quality of the manuscript.

This work was funded by the Deutsche Forschungsgemeinschaft (DFG, German Research Foundation) -- Project-ID 138713538 -- SFB 881 ('The Milky Way System', subproject A06). We thank Moritz Schmoll for his comments and contributions to finalise the paper.

PB and MI are grateful for the support from the special programme of the Polish Academy of Sciences and the US National Academy of Sciences under the long-term programme to support Ukrainian research teams grant No.~PAN.BFB.S.BWZ.329.022.2023.

PB and BS acknowledge the support of the Aerospace Committee of the Ministry of Digital Development, Innovations and Aerospace Industry of the Republic of Kazakhstan (Grant No.~BR20381077).

BS acknowledges financial support from the Science Committee of the Ministry of Science and Higher Education of the Republic of Kazakhstan (Grant No.~AP19677351 and AP13067834) and the Nazarbayev University Faculty Development Competitive Research Grant Programme (No.~11022021FD2912).

\end{acknowledgements}

\bibliographystyle{aa}
\bibliography{clusterbib}

\begin{appendix}

\onecolumn

\section{Angular momentum evolution of tail stars}\label{sec:app}

\begin{figure*}[h!]
\includegraphics[width=16.5cm]{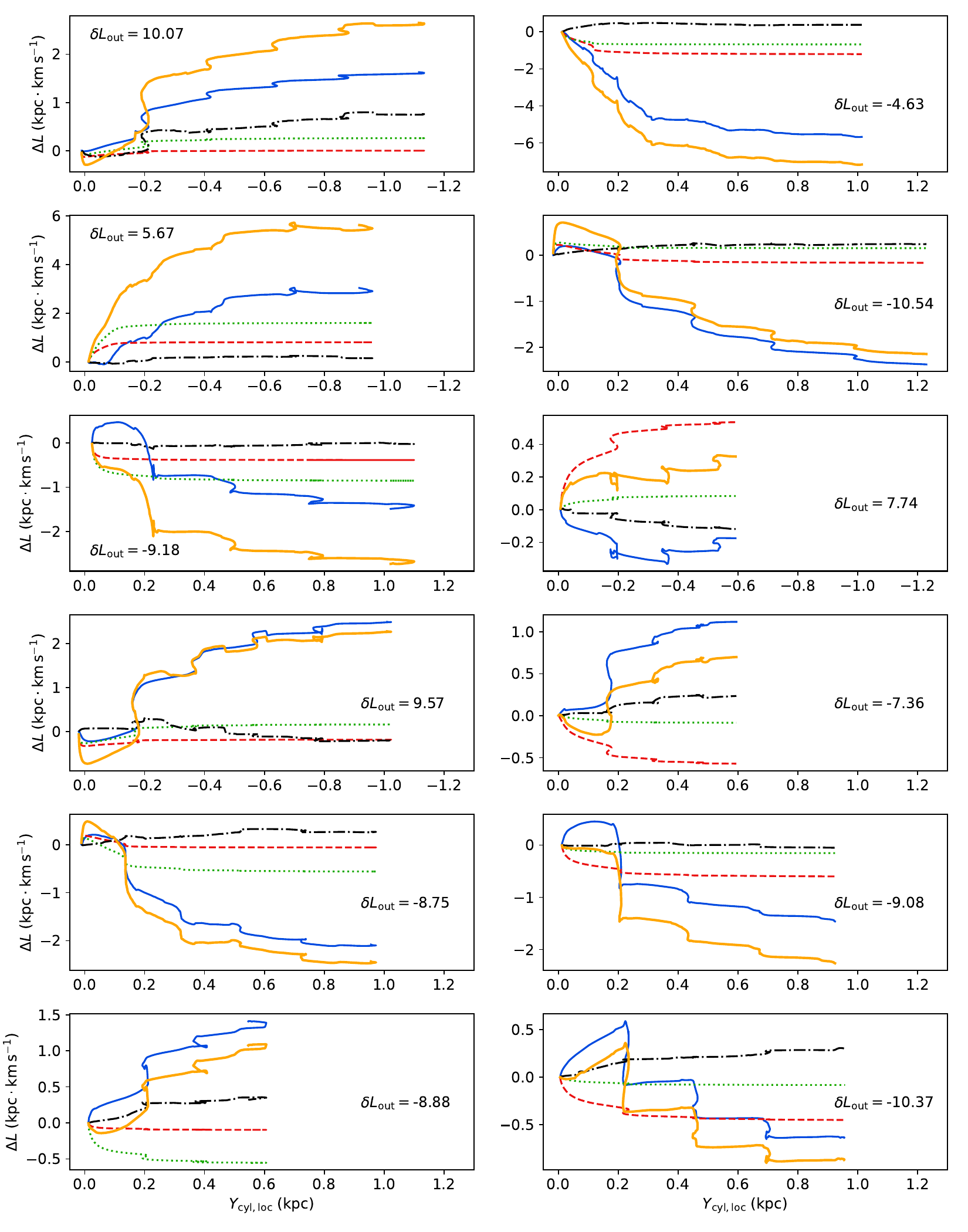}
\caption{Change of $\Delta L$ of stars over time along the tangential distance to the cluster centre split up into the subsystems. \textit{Left panels}: Stars from the BFSR. \textit{Right panels}: Stars of the COSR. Dashed red lines show the impact of the stars in the cluster. Thick blue lines (the stars in the tidal tails and dotted green lines) indicate the stars between $1r_\mathrm{J}$ radius and $2r_\mathrm{J}$. Dot-dashed black lines show the contribution to the angular momentum of the stars that were lost in the violent relaxation phase and thick amber lines (the total of all contributions). 
The direct measurement was omitted for clarity, as it consistently coincided with the total. 
Note that the $Y_\text{cyl,loc}$-axis for the stars in the trailing arm (negative values) is 
flipped. The unit, $\delta L_\mathrm{out}$, corresponds to $\mathrm{kpc\cdot km \, s^{-1}}$ and indicates the $z$-component of the star's angular momentum relative to the cluster centre at escape.}
\label{fig:appendix_angmom_plot}
\end{figure*}

\twocolumn


\end{appendix}

\end{document}